\documentclass[aps,reprint,superscriptaddress,showpacs,prb]{revtex4-1} 

\usepackage{amsmath}
\usepackage{amssymb}
\usepackage{graphicx}
\usepackage{bm}       
\usepackage{hyperref} 
\hypersetup{
     colorlinks=true,       
     linkcolor=blue,        
     citecolor=blue,        
     filecolor=blue,        
     urlcolor=cyan          
}
\usepackage[table]{xcolor}
\definecolor{lightGray}{gray}{0.95}
\definecolor{normalGray}{gray}{0.80}
\definecolor{darkGray}{gray}{0.65}

\newcommand{\nSPE}{\ensuremath{{\nu}}}
\newcommand{\Ni}{\ensuremath{{a}}}

\newcommand{\Lc}{\ensuremath{{\lambda_\mathrm{c}}}}
\newcommand{\Lcinf}{\ensuremath{{\lambda_\mathrm{c}^\infty}}}
\newcommand{\braket}[3]
{\ensuremath{\left\langle {#1} \left| {#2} \right| {#3}
    \right\rangle}}

\newcommand{\bra}[1]{\ensuremath{\langle #1 |}}
\newcommand{\ket}[1]{\ensuremath{| #1 \rangle}}

\renewcommand\Re{\operatorname{Re}}
\renewcommand\Im{\operatorname{Im}}
\renewcommand{\vec}[1]{\bm{ #1 }}
\DeclareMathOperator{\pf}{Pf}
\DeclareMathOperator{\sgn}{sgn}
\DeclareMathOperator{\tr}{tr}

\begin{document}
\title{Existence of zero-energy impurity states in different classes of topological insulators and superconductors and their relation to topological phase transitions}

\author{Lukas Kimme} \affiliation{Institut f\"ur Theoretische
Physik, Universit\"at Leipzig, D-04103 Leipzig, Germany}
\author{Timo Hyart} \affiliation{University of Jyv\"askyl\"a, Department of Physics and Nanoscience Center, P.O. Box 35 (YFL), FI-40014 University of Jyv\"askyl\"a, Finland}

\date{December 22, 2015}

\begin{abstract}
  We consider the effects of impurities on topological insulators and
  superconductors. We start by identifying the general conditions
  under which the eigenenergies of an arbitrary Hamiltonian $H$
  belonging to one of the Altland-Zirnbauer symmetry classes undergo a
  robust zero energy crossing as a function of an external parameter
  which can be, for example, the impurity strength. We define a
  generalized root of $\det H$, and use it to predict or rule out
  robust zero-energy crossings in all symmetry classes. We complement
  this result with an analysis based on almost degenerate perturbation
  theory, which allows a derivation of the asymptotic low-energy
  behavior of the ensemble averaged density of states
  $\rho \sim E^\alpha$ for all symmetry classes, and makes it
  transparent that the exponent $\alpha$ does not depend on the choice
  of the random matrix ensemble.  Finally, we show that a lattice of
  impurities can drive a topologically trivial system into a
  nontrivial phase, and in particular we demonstrate that impurity
  bands carrying extremely large Chern numbers can appear in different
  symmetry classes of two-dimensional topological insulators and
  superconductors. We use the generalized root of $\det H(\vec{k})$ to
  reveal a spiderweblike momentum space structure of the energy gap
  closings that separate the topologically distinct phases in
  $p_x+i p_y$ superconductors in the presence of an impurity lattice.
\end{abstract}

\maketitle
\section{Introduction}
\label{sec:Introduction}

Ever since the concept of topological order in condensed matter
systems picked up momentum during the last decade,\cite{Kitaev2001,
  Kane2005, Fu2007, Moore2007, Roy2009, Kitaev2009, Schnyder2008,
  Hasan2010, Qi2011} there has been the question as to which kind of
perturbations can cause topological phase transitions.  In this paper,
we investigate the possibility to utilize inhomogeneous perturbations,
like those caused by impurity lattices, for this purpose.  We point
out that such perturbations are often naturally present in candidate
materials for topological superconductors, since these are typically
unconventional superconductors obtained by doping Mott
insulators.\cite{Schnyder2008, Hyart2012, Okamoto2013, Scherer2014,
  You2012, BlackSchaffer2014} Moreover, high quality topological
insulators have recently been created by intentionally doping Si into
InAs/GaSb heterostructures.\cite{Du2015} Finally, a lattice of
magnetic atoms placed on top of a superconductor may be a route
towards the realization of a rich variety of topological
phases.\cite{Rontynen2015} In principle, there is also a relation to
the studies about so-called topological Anderson
insulators,\cite{Li2009, Groth2009} disorder-driven topological
superconductivity,\cite{Adagideli2014,Qin2015} and impurity bound
states as a signature of a topologically nontrivial
phase.\cite{Sau2013,Slager2015}

To intuitively understand why impurity lattices may be of interest for
the design of topologically nontrivial phases, a useful starting point
is to realize that nontrivial topological invariants are associated
with momentum space topological defects, \cite{Volovik-book} which can
be inserted into the system when the bulk energy gap closes. The
impurities reduce the translational symmetry of the system and give
rise to impurity bands, which can have complicated energy-momentum
dispersion relations described by long-range hopping terms in the
low-energy theory. This allows more freedom to deform the dispersion
in such a way that many band inversions appear, giving rise to
topological defects in momentum space. As a result these impurity
bands can carry very large topological invariants. Recently,
R\"ontynen and Ojanen theoretically demonstrated the possibility of
topological superconductivity with very large Chern numbers in 2D
ferromagnetic Shiba lattices using this kind of
approach. \cite{Rontynen2015}

In the first part of this paper (Secs.~\ref{sec:Roots} and
\ref{sec:DOS}), we investigate whether the eigenstates of an arbitrary
Hamiltonian $H_\lambda = H_0 + \lambda H_1$ belonging to one of the
ten Altland-Zirnbauer symmetry classes\cite{Altland1997,Zirnbauer1996}
undergo robust zero-energy crossings as a function of $\lambda$.  In
this regard, there are two main results: (i) We define the generalized
root $q(H)$ of $\det H$ in such a way that $|q(H)|^\nSPE = |\det H|$,
and $q(H)$ is a polynomial function of the matrix elements of the
Hamiltonian.  Here, $\nSPE$ is the symmetry-induced degeneracy of
eigenstates, when eigenstates at $\pm E$ are considered degenerate.
We show that $q(H)$ is real in classes A, AI, AII, BDI, and D, it is
complex in classes AIII, CI, and DIII, and it cannot be defined in
classes C and CII.  When $q$ is real, $\sgn q(H)$ equals the parity of
the topological invariant in $d=0$ dimensions.\cite{Kitaev2001,
  Kitaev2009, Fulga2012, Sau2013, Beenakker2013} In the presence of an
additional symmetry $[H,U]=0$, $\{C,U\}=0$, where $C$ is the chiral
symmetry operator, $q$ becomes real also in classes AIII, CI, and
DIII.\cite{Kimme2015} (ii) We use $q(H_\lambda)$ to predict
[$q(H_\lambda)$ is real] or to rule out [$q(H_\lambda)$ is complex]
robust zero-energy crossings in the respective symmetry classes. As a
side product of this analysis we also derive the important result that
if there are two parameters $\lambda_1$ and $\lambda_2$, robust zero
energy crossings can exist in the $(\lambda_1, \lambda_2)$-parameter
plane also when $q(H_{\lambda_1, \lambda_2})$ is complex.

We complement our exact algebraic formalism with almost degenerate
perturbation theory\cite{Marinescu1996} applied to an eigenstate
$\ket{\psi}$ of $H_0$ and its symmetry partner eigenstates (SPEs)
$\ket{T\psi}$, $\ket{P\psi}$, and $\ket{C\psi}$, where $T$, $P$, and
$C$ are the operators of time-reversal (TR), particle-hole (PH), and
chiral symmetry, respectively.  Using this approach in combination
with a scaling argument\cite{Altland1997,Haake2010}, we confirm the
predicted universality of the exponent
$\alpha$\cite{Altland1997,Ivanov2002,Beenakker2014} characterizing the
asymptotic behavior of the ensemble averaged density of states
$\rho \sim E^\alpha$ in the limit $E\to0$ for each of the ten symmetry
classes, respectively.  This approach makes transparent why the
exponent $\alpha$ does not depend on the choice of the random matrix
ensemble.

In the second part of this paper (Sec.~\ref{impurity-driven-trans}),
we study impurity-driven topological phase transitions. We discuss the
relation between the zero-energy crossings in $d=0$ and topological
phase transitions in $d>0$ for systems where the topological invariant
${\cal Q}$ is changed by band inversion at the high-symmetry points
$\vec{\Gamma} = -\vec{\Gamma} + \vec{G}$ of the Brillouin zone
($\vec{G}$ is a reciprocal lattice vector). Our approach is to first
consider a generic translationally invariant Hamiltonian $H_0$ and a
local perturbation $\lambda H_1$ causing a zero-energy crossing at
$\lambda = \Lcinf$.  The robust zero-energy crossings are then
utilized to provide an estimate for the impurity strengths necessary
to drive topological phase transitions in the presence of an impurity
lattice. For this purpose, we analyze $q(H_\lambda(\vec{\Gamma}))$,
where $H_\lambda = H_0 + \lambda H_1'$ and $H_1'$ describes the
impurity lattice formed by periodically continuing $H_1$ with a
lattice constant $\Ni$.  For each $\vec{\Gamma}$, a critical value
$\Lc(\vec{\Gamma})$ exists, and they all coincide
$\Lc(\vec{\Gamma}) = \Lcinf$ in the limit $\Ni/\xi^\infty\to\infty$,
where $\xi^\infty$ is the decay length of the zero-energy eigenstate
wave function of $H_0 + \Lcinf H_1$.  However, at finite $\Ni$, the
$\Lc(\vec{\Gamma})$ being unequal give rise to a range of values
$\lambda$ where $H_\lambda$ is nontrivial even though $H_0$ is
trivial.  We illustrate this general behavior analytically in the
context of the 1D Kitaev model\cite{Kitaev2001} (equivalently the SSH
model\cite{Su1979}).

The closing of the energy gap can also occur away from the
high-symmetry points of the Brillouin zone. In this case, one can
consider $\vec{k}$ as a parameter and use the zero dimensional
$q(H_\lambda(\vec{k}))$ to analyze the phase transitions. The main
difference is that the zero-dimensional Hamiltonian
$H_\lambda(\vec{k})$, where $\vec{k}$ is considered to be a fixed
parameter, usually belongs to a different symmetry class than the full
higher dimensional Hamiltonian containing all the momentum blocks.  We
have studied the impurity lattices in several two-dimensional models,
which show this kind of gap closings away from the high-symmetry
points.  From these closings arises a rich variety of topological
phases.  In particular, we demonstrate that similarly as in the case
of ferromagnetic Shiba lattices,\cite{Rontynen2015} it is possible to
obtain large Chern numbers also in $p_x+ip_y$ superconductors, but in
this case one only needs nonmagnetic impurities instead of magnetic
atoms. Furthermore, we use $q(H_\lambda(\vec{k}))$ to reveal a
spiderweblike momentum space structure of the energy gap closings that
separate the topologically distinct phases in $p_x+i p_y$
superconductors in the presence of an impurity lattice.  Both the
Shiba lattice\cite{Rontynen2015} and the $p_x+i p_y$ superconductor
models belong to class D in the classification table for topological
insulators and superconductors.\cite{Schnyder2008} However, the idea
of using an impurity lattice to design topologically nontrivial phases
with large Chern numbers is more general. For this purpose we
demonstrate that an impurity lattice can give rise to large Chern
numbers also in Chern insulators belonging to class A in the
classification table.

\begin{table*}
  \caption{The first four columns define the symmetry classes. 
    The absence of symmetry is denoted as "0" and if TR or PH
    symmetries are present the values of $\epsilon_T = \pm 1$
    and $\epsilon_P = \pm 1$ are given. $\nSPE$ is the number
    of SPEs.  $q(H)$ is the generalized $\nSPE$th root of
    $\det H$, satisfying $|q(H)|^\nSPE = |\det H|$.  $q(H)$
    takes values in ${\cal W}$.  ${\cal Q}$ is the topological invariant
    in $d=0$.  $\alpha$ is the exponent characterizing the asymptotic
    behavior of the ensemble averaged density of states $\rho(E)\sim|E|^\alpha$ in the
    limit $E\to 0$.  In classes with chiral
    symmetry, in the presence of an additional symmetry $[H,U]=0$,
    $\{C,U\}=0$, the exponent $\alpha$ is lowered by one
    $\alpha_U = \alpha-1$.  In classes AIII, CI, and DIII, a symmetry
    $U$ ensures that $q(H)$ is real up to a phase $e^{i\varphi(U)}$ which
    only depends on $U$. By taking this phase factor into account in
    the definition of $q$ (denoted $q_U$ in the text),  ${\cal W}$ is
    reduced to ${\cal W}_U$. Moreover, $U$ enables the definition of
    a vector ${\cal Q}_U$ of topological invariants in these classes.}
  \label{tab:results}
  \begin{ruledtabular}
  \begin{tabular}{lrrcccrccrcc}
     & $T^2$ & $P^2$ & $C^2$ & $\nSPE$ & $q(H)$           & ${\cal W}$                         & ${\cal Q}$     & $\alpha$ & ${\cal W}_U$                       & ${\cal Q}_U$                     & $\alpha_U$\\\hline
\colorbox{black!20!white}{AIII} & $0$   & $0$   & $1$   & $2$     & $\det D$         & $\mathbb{C}$                       & $0$            & $1$      & $\mathbb{R}$        & $\mathbb{Z}^n$                   & $0$\\
\colorbox{black!5!white}{A}    & $0$   & $0$   & $0$   & $1$     & $\det H$         & $\mathbb{R}$                       & $\mathbb{Z}$   & $0$      & $-$                                & $-$                              & $-$\\\hline
\colorbox{black!20!white}{CI}   & $+1$  & $-1$  & $1$   & $2$     & $\det D$         & $\mathbb{C}$                       & $0$            & $1$      & $\mathbb{R}$        & $\mathbb{Z}^n$                   & $0$\\
\colorbox{black!5!white}{AI}   & $+1$  & $0$   & $0$   & $1$     & $\det H$         & $\mathbb{R}$                       & $\mathbb{Z}$   & $0$      & $-$                                & $-$                              & $-$\\
\colorbox{black!5!white}{BDI}  & $+1$  & $+1$  & $1$   & $2$     & $\sqrt{\frac{(-1)^N}{\det{\cal P}}}\pf(H{\cal P})$ & $\mathbb{R}$ & $\mathbb{Z}_2$ & $0$      & $\mathbb{R}$ & $\mathbb{Z}^n$                   & $0$\\
\colorbox{black!5!white}{D}    & $0$   & $+1$  & $0$   & $2$     & $\sqrt{\frac{(-1)^N}{\det{\cal P}}}\pf(H{\cal P})$ & $\mathbb{R}$ & $\mathbb{Z}_2$ & $0$      & $-$                                & $-$                              & $-$\\
\colorbox{black!20!white}{DIII} & $-1$  & $+1$  & $1$   & $4$     & $\pf(D)$         & $\mathbb{C}$                       & $0$            & $1$      & $\mathbb{R}$        & $(\mathbb{Z}^{n_1}, \mathbb{Z}_2^{n_2})$ & $0$\\
\colorbox{black!5!white}{AII}  & $-1$  & $0$   & $0$   & $2$     & $\sqrt{\frac{1}{\det{\cal T}}}\pf(H{\cal T})$ & $\mathbb{R}$ & $\mathbb{Z}$   & $0$      & $-$                                & $-$                              & $-$\\
\colorbox{black!35!white}{CII}  & $-1$  & $-1$  & $1$   & $4$     & $-$              & $-$                                & $0$            & $3$      & $-$                                & $-$                              & $2$\\
\colorbox{black!35!white}{C}    & $0$   & $-1$  & $0$   & $2$     & $-$              & $-$                                & $0$            & $2$      & $-$                                & $-$                              & $-$
  \end{tabular}
  \end{ruledtabular}
\end{table*}

\section{Generalized roots of $\det H$ and relation to zero-energy
  crossings}
\label{sec:Roots}
There are three types of symmetries -- chiral, TR, and PH symmetry --
depending on the presence or absence of which a given Hamiltonian
matrix $H$ is said to belong to one of ten symmetry
classes:\cite{Altland1997,Zirnbauer1996}
\begin{subequations}
\label{eq:symmetries}
 \begin{align}
   \mathrm{chiral:}& &\{H,C\} &= 0, & C^2 &= 1 \label{eq:chiral}\\
   \mathrm{TR:}& &[H,T] &= 0, & T^2 &= \epsilon_T \label{eq:TRS}\\
   \mathrm{PH:}& &\{H,P\} &= 0, & P^2 &= \epsilon_P \label{eq:PHS}
 \end{align}
\end{subequations}
where $T={\cal T}K$, $P={\cal P}K$, $K$ is the operator of complex
conjugation, and $C$, ${\cal T}$, and ${\cal P}$ are unitary matrices.
TR and PH symmetry are characterized by the sign $\epsilon_T = \pm 1$
and $\epsilon_P = \pm 1$, respectively.  For a list of symmetry
classes (determined by the absence ``$0$'' or the presence of various
symmetries with $\epsilon_T = \pm 1$, $\epsilon_P = \pm 1$) and an
overview of results, see Table~\ref{tab:results}.  In the presence of
a symmetry $C$, $T$, or $P$, one finds for each eigenstate
$\ket{\psi}$ of $H$ so-called SPEs $\ket{C\psi}$, $\ket{T\psi}$, or
$\ket{P\psi}$, respectively.  Equations~\eqref{eq:symmetries} imply
that $\ket{T\psi}$ ($\ket{C\psi}$ and $\ket{P\psi}$) is an eigenstate
of $H$ with the same (with opposite) energy as $\ket{\psi}$.  However,
TR symmetry with $T^2=1$ does not guarantee the existence of an
orthogonal symmetry partner eigenstate (SPE), because the Kramers
theorem applies only to the case $T^2=-1$.  Therefore, in the present
context such a symmetry should rather be understood as a constraint on
$H$.  The number of SPEs $\nu$ for all ten symmetry classes are shown
in Table~\ref{tab:results}.

The determinant of a matrix equals the product of its eigenvalues
$\det H = \prod_i E_i$.  Hence, $\det H$ allows us to analyze
zero-energy crossings and $d=0$ topological transitions.  We show that
in all symmetry classes, except for C and CII, one can explicitly
define a function $q(H)$ with two key properties:
\begin{itemize}
\item[(i)] $q(H)$ is a (complex valued) polynomial expression in the
  matrix elements $H_{ij}$ of the Hamiltonian.
\item[(ii)] $q(H)$ satisfies $|\det H| = |q(H)|^\nSPE$, where $\nSPE$
  is the number of SPEs in the respective symmetry class.
\end{itemize}
Property (ii) defines in which sense it is justified to call $q(H)$
the (generalized) $\nSPE$th root of $\det H$.  However, property (i)
constitutes the important difference between $q(H)$ and
$\sqrt[\nSPE]{\det H}$.  Most importantly, $q(H)$ shares with $\det H$
the property that it vanishes if and only if the Hamiltonian has a
zero eigenvalue. Before defining $q$ in Sec.~\ref{sec:definingq}, we
now explain its relation to the $d=0$ topological invariants.  Its
relation to zero-energy crossings is detailed in
Sec.~\ref{sec:qZeroEnergyCrossings}.

From $|\det H| = |q(H)|^\nSPE$ and the possibility to calculate
$\det H=\prod_i E_i$ as the product of the Hamiltonian's eigenvalues
follows
\begin{equation}
  q(H) = \sideset{}{'}\prod_i E_i,
\end{equation}
when $q$ is real.  Here, the prime indicates that only one member of
the set of \nSPE\ SPEs contributes a factor $E_i$.  In principle, $q$
is only defined up to a sign, but once the sign is chosen, it remains
well defined when changing $H$ adiabatically.  In consequence, we find
that in those symmetry classes where the $d=0$ invariant ${\cal Q}$
exists, $q$ changes sign when ${\cal Q}$ changes by $\pm 1$.  On the
other hand, in the presence of an additional symmetry $U$ in classes
AIII, CI, and DIII, $H$ is block diagonal with blocks $H_{\pm i}$
($i=1,2,...,n$), and if $\{C,U\}=0$ an invariant
${\cal Q}_{H_{\pm i}}$ exists for each block;
cf. Appendix~\ref{sec:Implications}.  Then, $q$ changes sign when one
of the ${\cal Q}_{H_{+i}}$ changes by $\pm 1$.  Formally, this
corresponds to
\begin{equation}
  \sgn q(H) =  \begin{cases}
    (-1)^{\cal Q} & \text{A, AI, AII, BDI, D}\\
    \prod_{i=1}^n (-1)^{{\cal Q}_{H_i}} & \text{AIII+$U$, CI+$U$, DIII+$U$}
  \end{cases}
\end{equation}
where X+$U$ means ``for Hamiltonians from symmetry class X with the
additional symmetry $U$, $\{C,U\}=0$''.  For examples of unitary
symmetries in the context of topological order see
Refs.~\onlinecite{Kimme2015, Woelms2014, Koshino2014, Fu2011,
  Fang2012, Slager2013, Ueno2013, Zhang2013, Chiu2013}.

\subsection{Defining generalized roots of $\det H$}
\label{sec:definingq}

\paragraph*{Classes A and AI.} Since SPEs are absent, we can define
$q(H) = \det H$.

\paragraph*{Classes BDI and D.} Hamiltonians in these symmetry classes
obey PH symmetry with $P^2 = +1$.  In accordance with
Refs.~\onlinecite{Kitaev2001, Ghosh2010, Fulga2012}, we define the
real polynomial $q(H) = \sqrt{(-1)^N/\det{\cal P}} \pf(H{\cal P})$,
where $H$ is a $2N \times 2N$ matrix.  Using
$({\cal P}K)^2 = {\cal P}{\cal P}^\dagger = +1$ and Eq.~\eqref{eq:PHS}
it follows that ${\cal P} = {\cal P}^T$ and
$H{\cal P} = -(H{\cal P})^T$. Thus $\pf(H{\cal P})$ is well defined.
From the general property of the Pfaffian $\pf(A)^2 = \det(A)$, we
then infer $|q(H)|^2 = |\det H|$.

\paragraph*{Class AII.} Hamiltonians in AII obey TR symmetry with
$T^2 = -1$.  We proceed analogously to the derivation of $q(H)$ for
classes BDI and D and thus define the real polynomial
$q(H) = \sqrt{1/\det{\cal T}} \pf(H{\cal T})$.

\paragraph*{Classes AIII and CI.} Hamiltonians in these classes have a
chiral symmetry.  We focus on the case where
$\tr C = 0$.\cite{footnote:ChiralSymmetry} This allows us to choose a
basis such that
\begin{subequations}
\begin{align}
  C &= \begin{pmatrix}
    \bm{1}_N & \bm{0}_N \\ \bm{0}_N & -\bm{1}_N
  \end{pmatrix}, \label{eq:Cdiagonal} \\
  H &= \begin{pmatrix}
    \bm{0}_N & D \\ D^\dagger & \bm{0}_N \label{eq:HamiltonianBlockoffdiagonal}
  \end{pmatrix}.
\end{align}
\end{subequations}
Since $|\det H| =  |\det D|^2$ we can straightforwardly define the
complex polynomial $q(H) = \det D$.

Chiral symmetry alone does not put any restriction on $D$.  Albeit TR
symmetry restricts $D$ in class CI, the phase of $\det D$ does depend
on the parameters of the Hamiltonian in both classes, CI and AIII.  In
the following, we show that the phase of $\det D$ becomes independent
of the Hamiltonian's parameters if there is a symmetry $U$ satisfying
$[H,U] = \{C,U\} = 0$.  In the basis of Eq.~\eqref{eq:Cdiagonal},
$\{C,U\} = 0$ implies
\begin{equation}
  U = \begin{pmatrix}
    \bm{0}_N & U_1 \\ U_2 & \bm{0}_N
  \end{pmatrix},
\end{equation}
with unitary matrices $U_1$ and $U_2$.  The symmetry $[H,U]=0$ implies
$U_1 D^\dagger = DU_2$.  This can be used to show
$(\det D)^* = \det(U_1^\dagger U_2) \det D$, which means that the
complex phase of $q(H)$ is independent of $H$ in the presence of the
additional symmetry $U$.  Hence, we define the real polynomial
$q_U(H) = \sqrt{\det(U_1^\dagger U_2)} \det D$.

\paragraph*{Class DIII.} A polynomial expression for the fourth root
of $\det H$ is a central result of Ref.~\onlinecite{Kimme2015}.  For
completeness, we briefly summarize the derivation here.  Due to chiral
symmetry, $H$ is of off-diagonal form, cf.
Eq.~\eqref{eq:HamiltonianBlockoffdiagonal}.  TR symmetry with
$T^2 = -1$ renders $D = -D^T$ antisymmetric.  This enables the
definition $q(H) = \pf D$.  Also in class DIII, the complex phase of
$q(H)$ loses its dependence on the Hamiltonian's parameters in the
presence of a symmetry $[H,U] = \{C,U\} = 0$. Then, the real
polynomial $q_U(H) = \sqrt{(-1)^N/\det W}\pf D$ can be defined, where
$W$ is a block of $U = \left(\begin{smallmatrix} 0 & W \\ W^* & 0
  \end{smallmatrix}\right)$
in the basis where $C$ is diagonal, cf. Eq.~\eqref{eq:Cdiagonal}.

\paragraph*{Classes C and CII.} In these classes, it is not possible
to define $q(H)$ as a polynomial in the matrix elements of $H$.  For a
justification, see Appendix~\ref{sec:classesCandCII}.

\subsection{Relation to zero-energy crossings}
\label{sec:qZeroEnergyCrossings}

\begin{figure}
  \includegraphics[width=\columnwidth]{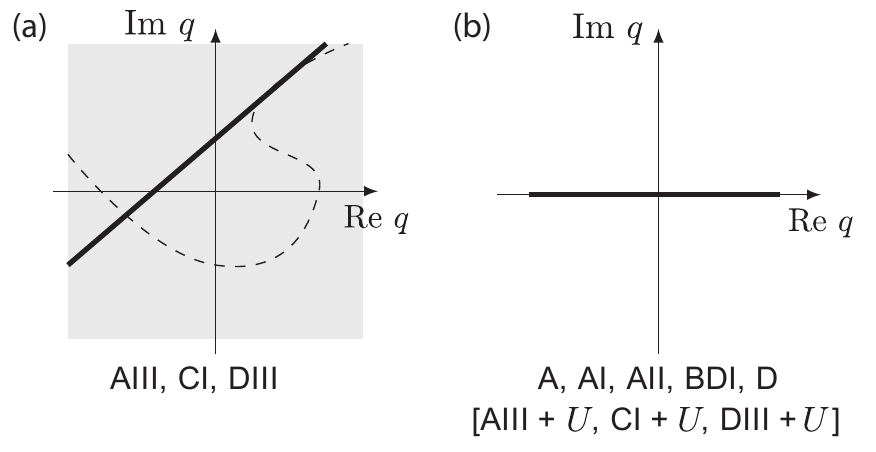}
  \caption{\label{fig:qH} Illustration of the real and imaginary parts
    of $q(H_\lambda)$ as a function of an external parameter
    $\lambda$. (a) In classes AIII, CI, and DIII, $q(H_\lambda)$ takes
    complex values and generically does not become zero, independently
    of whether a parameter is considered on which $q$ depends linearly
    (solid line) or with some higher power (dashed line).  (b) In
    classes A, AI, AII, BDI, and D [AIII, CI, and DIII with additional
    symmetry $U$ which anticommutes with $C$] $q(H_\lambda)$
    [$q_U(H_\lambda)$] is always real.  In these cases the change of
    sign of $q(H_\lambda)$ as a function of $\lambda$ guarantees the
    existence of robust zero-energy crossing at some value
    $\lambda=\lambda_c$. }
\end{figure}

Since $|q(H)|^\nSPE = |\det H|$, the zeros of $q(H)$ coincide with the
zeros of $\det H$ and the existence of zero-energy eigenstates of $H$.
For predicting zero-energy crossings, $q(H)$ is more useful than
$\det H$ because (i) it is a complex number in classes AIII, CI, and
DIII and thus has a phase degree of freedom and (ii) it accounts for
symmetries and thus is a polynomial of lower degree than $\det H$. In
Sec.~\ref{sec:DOS} we supplement the following abstract discussion
with additional more intuitive arguments.

In Ref.~\onlinecite{Kimme2015}, the possibility to predict zero-energy
crossings using $q(H)$ was illustrated for a Hamiltonian
$H_\lambda = H_0 + \lambda H_1$, where $H_0$ is from DIII and
$\lambda H_1$ is an on-site potential with impurity strength
$\lambda$.  While $\det H_\lambda$ is a real fourth order polynomial
in $\lambda$, $q(H_\lambda)$ is a complex first order
polynomial. Consequently, $q(H_\lambda)$ as a function of $\lambda$
takes values on a straight line in the complex plane which generically
does not contain the origin, so that zero-energy eigenstates will not
occur for any $\lambda$; cf. solid line in Fig.~\ref{fig:qH}(a).  More
generally, $q(H)$ describes a continuous curve in the complex plane
under arbitrary continuous parameter changes of $H$ and will not go
through the origin, so that it will not give rise to zero-energy
states without fine tuning; cf. dashed line in Fig.~\ref{fig:qH}(a).
As a result, in classes AIII, CI, and DIII, where $q(H)\in\mathbb{C}$,
changing a single parameter in $H$ will not induce robust zero-energy
crossings.

\begin{figure}
  \includegraphics[width=.8\columnwidth]{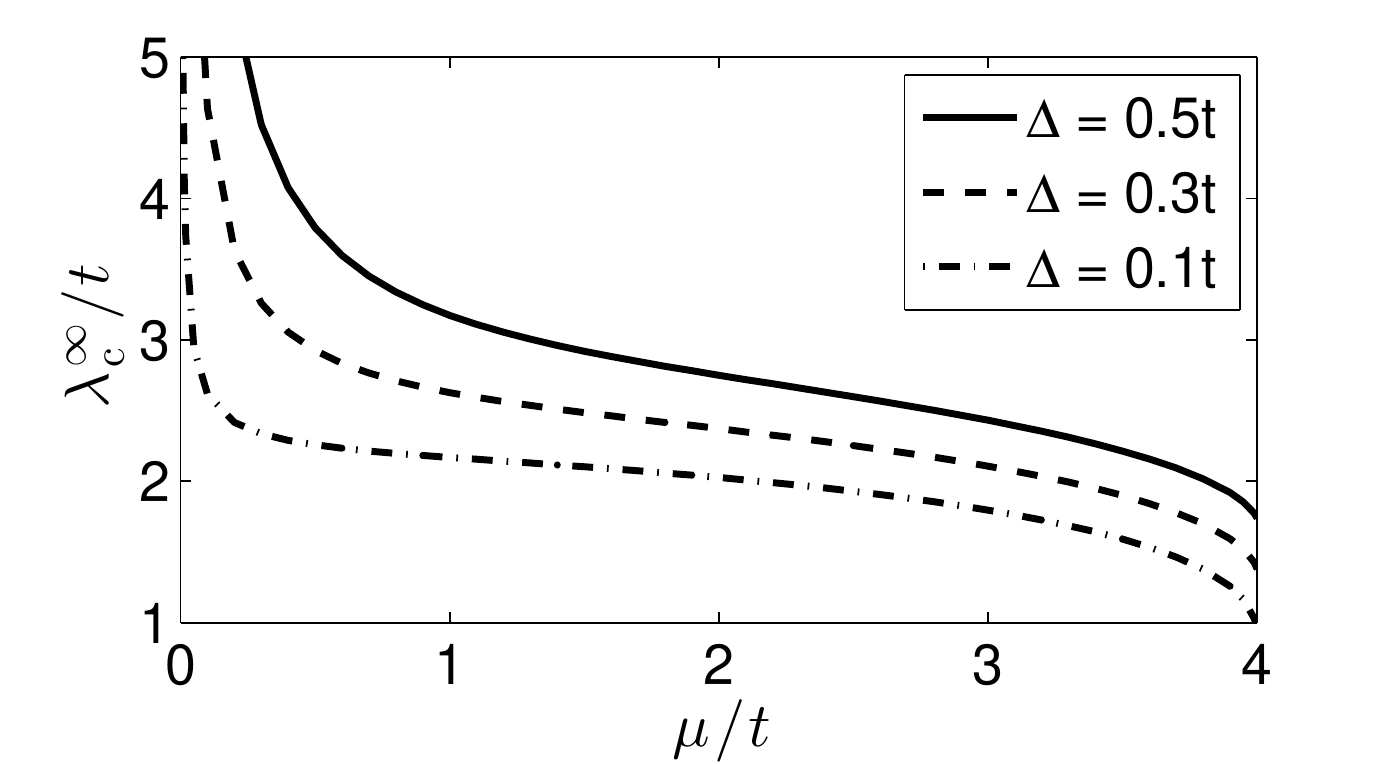}
  \caption{\label{fig:ucinfinite} The critical impurity strength
    \Lcinf\ of an on-site potential impurity
    $\lambda H_1 = \lambda c_0^\dagger c_0$ in a $p_x+ip_y$
    superconductor depends on the chemical potential $\mu$.  Hence,
    zero-energy crossings appear not only by tuning $\lambda$, but
    also by tuning $\mu$.}
\end{figure}

\begin{figure}
  \includegraphics[width=.8\columnwidth]{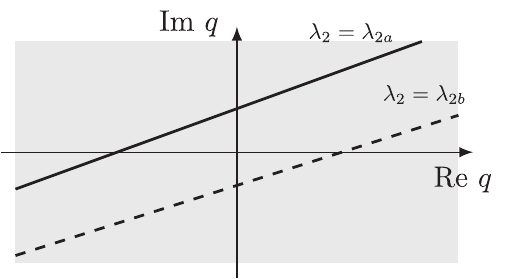}
  \caption{\label{fig:crossinginplane} Illustration of the real and
    imaginary parts of $q(H_{\lambda_1, \lambda_2})$ as a function of
    an external parameter $\lambda_1$ for two fixed values of another
    external parameter $\lambda_2= \lambda_{2a}$ and
    $\lambda_2=\lambda_{2b}$ in classes where $q(H)$ takes complex
    values. If the two curves are on opposite sides of the origin as
    illustrated in the figure and assuming that for
    $\lambda_2 \in [\lambda_{2a}, \lambda_{2b}]$ $\Re q$ (or $\Im q$)
    is a continuously increasing/decreasing function of $\lambda$ with
    $\min [\Re q]<0$ and $\max [\Re q]>0$ ($\min [\Im q]<0$ and
    $\max [\Im q]>0$), these curves cannot be smoothly deformed into
    each other except if $q(H_{\lambda_1, \lambda_2})=0$ at a
    particular point $\lambda_{1}=\lambda_{1c}$ and
    $\lambda_2=\lambda_{2c}$. Therefore there exists a robust
    zero-energy crossing within the
    $(\lambda_{1}, \lambda_{2})$-parameter plane.}
\end{figure}

\begin{figure*}
  \includegraphics[width=\textwidth]{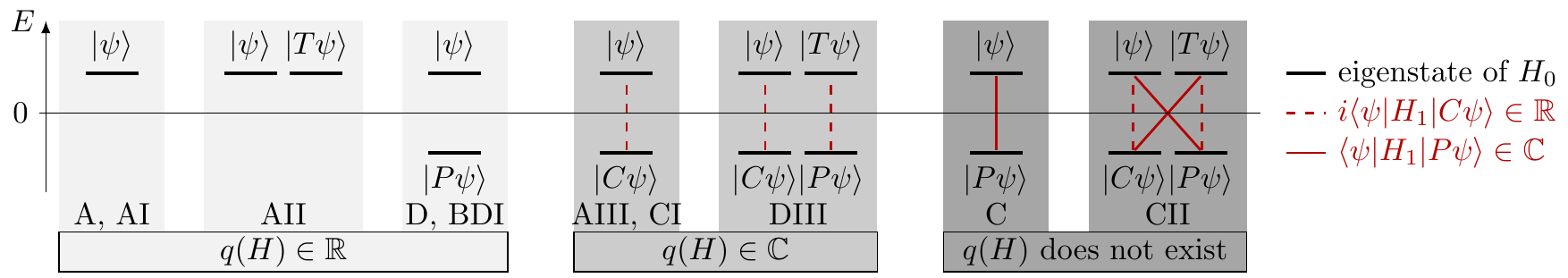}
  \caption{\label{fig:SPE} For each symmetry class, an eigenstate
    $\ket{\psi}$ of $H_0$ and its SPEs are shown.  A perturbation
    $H_1$ from the same class as $H_0$ couples SPEs with opposite
    energies as indicated in the figure.  The possible couplings to
    chiral (PH) partners are shown by dashed (solid) red lines.  The
    coupling to the chiral SPE vanishes when the Hamiltonian obeys an
    additional symmetry $U$ with $\{C,U\}=0$.}
\end{figure*}

Consider now the case where $q(H)$ is real.  Focusing on
$H_\lambda = H_0 + \lambda H_1$, where $H_1$ is an arbitrary local
perturbation, we find $q(H_\lambda) = \sum_{i=0}^n b_i\lambda^i$,
where $b_i\in\mathbb{R}$ and $n$ depends on $H_1$.  Clearly, when $n$
is odd, $q(H_\lambda)$ crosses zero at least once upon tuning
$\lambda$ from $-\infty$ to $+\infty$ [Fig.~\ref{fig:qH}(b)].  For
even $n$, an even number of zero-energy crossings will occur,
including the possibility that $q$ does not cross zero.

Some remarks are in order.  (i) Although the zero-energy crossings are
most easily analyzed as a function of $\lambda$, they appear also as a
function of $H_0$ bulk parameters like the chemical potential.  This
is illustrated in Fig.~\ref{fig:ucinfinite}, where we plot the
critical impurity strength $\Lcinf$ corresponding to a zero-energy
crossing for an impurity $\lambda H_1 = \lambda c_0^\dagger c_0$,
where $c_i$ annihilates a spinless electron on site $i$, in an
infinite $p_x+ip_y$ superconductor.  The unperturbed Hamiltonian in
momentum space reads
\begin{equation}
  H_0(\vec{k}) = \frac{1}{2}\sum_{\vec{k}} (c_{\vec{k}}^\dagger, c_{-\vec{k}})
  \begin{pmatrix}
    h(\vec{k}) & \Delta(\vec{k}) \\
    \Delta(\vec{k})^* & -h(\vec{k})
  \end{pmatrix}
  \begin{pmatrix}
    c_{\vec{k}} \\
    c_{-\vec{k}}^\dagger \\
  \end{pmatrix}
  \label{eq:pxipy}
\end{equation}
with
\begin{subequations}
  \begin{align}
    h(\vec{k}) &=  - 2t(\cos k_x + \cos k_y) -\mu, \\
    \Delta(\vec{k}) &= \Delta(\sin k_x+i\sin k_y).
  \end{align}
\end{subequations}
Here $t$ and $\Delta$ parametrize the nearest neighbor hopping and
superconducting pairing, respectively, and $\mu$ is the chemical
potential.  \Lcinf\ was determined by performing a $T$-matrix
calculation.\cite{Balatsky2009} (ii) In classes AIII, CI, and DIII the
presence of a symmetry $U$ is not a necessary but a sufficient
condition to render the phase of $q(H)$ independent of
$H$.\cite{Kimme2015} In fact, contrary to the set of equations
$[H,U]=0$, the necessary and sufficient condition
$\Im q = \tan \varphi \Re q$ is just a single equation.  (iii) The
argument for the existence of zero-energy crossings when $q$ is real
can be generalized to interacting models where $H_0$ needs to be
calculated self-consistently when the external parameter $\lambda$ is
varied. Assuming that the coefficients $b_i$ become continuous
functions of $\lambda$ and there is no spontaneous symmetry breaking,
the change of sign of $q(H_\lambda)$ still guarantees the existence of
a robust zero-energy crossing. (iv) If $q(H_\lambda) \in \mathbb{C}$
it can still be used to predict robust zero-energy crossings. However,
in this case one needs at least two external parameters
$(\lambda_{1}, \lambda_{2})$, and the robust crossings can only be
guaranteed to occur at particular critical points
$(\lambda_{1c}, \lambda_{2c})$ within a parameter plane. Namely,
consider the real and imaginary parts of $q(H_{\lambda_1, \lambda_2})$
as a function of an external parameter $\lambda_1$ for two fixed
values of another external parameter $\lambda_2= \lambda_{2a}$ and
$\lambda_2=\lambda_{2b}$. If the two curves are on opposite sides of
the origin as illustrated in Fig.~\ref{fig:crossinginplane}, under
certain assumptions they cannot be smoothly deformed into each other
unless $q(H_{\lambda_1, \lambda_2})=0$ at a particular point
$\lambda_{1}=\lambda_{1c}$ and $\lambda_2=\lambda_{2c}$. Therefore
this kind of situation typically leads to the existence of a robust
zero-energy crossing at a specific point within the
$(\lambda_{1}, \lambda_{2})$-parameter plane.

\section{Level repulsion and density of states in random matrix
  ensembles}
\label{sec:DOS}
There are relations between $q(H)$ and the allowed couplings of SPEs
as well as the density of states in random matrix ensembles.  These
relations are the subject of this section.

We focus on the eigenstate $\ket{\psi}$ of a Hamiltonian $H_0$ which,
together with its SPEs, is the one with energy closest to zero.
Subsequently, we study how the energy corresponding to this set of
SPEs evolves as the strength of a perturbation $H_1$ from the same
symmetry class as $H_0$ is tuned.  Therefore, we employ almost
degenerate perturbation theory\cite{Marinescu1996} in the subspace
spanned by these SPEs.  Moreover, we consider ensembles of random
matrices from any of the ten symmetry classes.  Then, by adopting a
scaling argument from Ref.~\onlinecite{Haake2010}, we use our
perturbative results to derive the universal asymptotic behavior of
the respective ensemble averaged density of states in the zero-energy
limit.

As an immediate consequence of the defining symmetry properties
Eqs.~\eqref{eq:symmetries}, the matrix elements describing the
coupling of SPEs are restricted as follows:
\begin{subequations}
\label{eq:SymmetryConstraints}
 \begin{align}
   \braket{\psi}{H_1}{C\psi} &= - \braket{\psi}{H_1}{C\psi}^*, \label{eq:couplingChiralSPE}\\
   \braket{\psi}{H_1}{T\psi} &= \epsilon_T \braket{\psi}{H_1}{T\psi}, \\
   \braket{\psi}{H_1}{P\psi} &=  -\epsilon_P \braket{\psi}{H_1}{P\psi}. \label{eq:couplingPHSPE}
 \end{align}
\end{subequations}
The possible couplings, restricted by these conditions, are depicted
in Fig.~\ref{fig:SPE} for all ten symmetry classes.  Assuming that the
energy $\epsilon$ of $\ket{\psi}$ is sufficiently close to zero, the
effect of a generic perturbation $H_1$ from the same symmetry class as
$H_0$ on $\ket{\psi}$ and its SPEs can be calculated via almost
degenerate perturbation theory.\cite{Marinescu1996} The energies of
the set of almost degenerate states in the presence of the
perturbation is obtained from the eigenvalues of
$\bar{H} = \Pi_{\ket{\psi}}(H_0+H_1)\Pi_{\ket{\psi}}$, where
$\Pi_{\ket{\psi}}$ projects onto the subspace spanned by $\ket{\psi}$
and its SPEs.

As an example we consider symmetry class AIII, where the matrix
elements for the projected perturbation read
$\bra{\psi}H_1\ket{\psi} = -\bra{C\psi}H_1\ket{C\psi} = \epsilon_1$,
and $\bra{\psi}H_1\ket{C\psi} = i\epsilon_2$;
$\epsilon_1, \epsilon_2 \in\mathbb{R}$.  Hence, we find in the basis
$(\ket{\psi}, \ket{C\psi})$
\begin{align}
  \bar{H}^\mathrm{AIII} &= 
  \begin{pmatrix}
    \epsilon+\epsilon_1 & i\epsilon_2 \\
    -i\epsilon_2 & -\epsilon-\epsilon_1
  \end{pmatrix},\\
  E_\pm^\mathrm{AIII} &= \pm\sqrt{(\epsilon+\epsilon_1)^2 + \epsilon_2^2}\ .
\end{align}
Because $\ket{\psi}$ couples to its chiral partner, one generically
expects an avoided level crossing upon tuning the perturbation.

Exploiting the symmetry constraints
[Eqs.~\eqref{eq:SymmetryConstraints}], one can derive analogous
results for all ten symmetry classes.  The corresponding energies are
always of the form
$E_+ = \sqrt{(\epsilon+\epsilon_1)^2 + \epsilon_2^2 + \dots +
  \epsilon_{\alpha+1}^2}$
with $\alpha = 0,1,2,$ or $3$ depending on the symmetry class
considered. The corresponding values of $\alpha$ for all ten symmetry
classes are shown in Table~\ref{tab:results}, and as we show below
$\alpha$ is the universal exponent determining the low-energy
asymptotic behavior of the ensemble averaged density of states.
$\alpha=0$ is to be understood as $E_+=\epsilon+\epsilon_1$.  $E_+$ is
twice degenerate in classes AII, DIII, and CII.  In those classes with
a PH symmetric spectrum one also gets $E_- = -E_+$.

We now consider arbitrary random matrix ensembles from any of the ten
symmetry classes.  Our perturbative results can then be used to
calculate the universal asymptotic behavior of the respective ensemble
averaged density of states $\rho(E)$ in the limit
$E\to 0$.\cite{Haake2010} To this end, we consider the sum $H_0 + H_1$
to be a realization of a random matrix from the respective ensemble.
The $\epsilon_i$ are then random variables with the corresponding
probability distribution $W(\vec{\epsilon})$,
$\vec{\epsilon}=(\epsilon_1, \dots, \epsilon_{\alpha+1})$, which we do
not have to know in detail.  $\epsilon$ is not independent of
$\epsilon_1$ and thus has been absorbed in $\epsilon_1$.  The ensemble
averaged density of states reads
\begin{equation}
  \rho(E) = 
  \int\! \mathrm{d}^{(\alpha+1)}\epsilon\
  W(\vec{\epsilon})\delta (|E| - \sqrt{\vec{\epsilon}^2} )\ .
\end{equation}
Upon rescaling $\vec{\epsilon} \mapsto |E| \vec{\epsilon}$, we obtain
$\rho(E) = |E|^\alpha \int \! \mathrm{d}^{(\alpha+1)}\epsilon\,
W(|E|\vec{\epsilon})\delta(1 - \sqrt{\vec{\epsilon}^2} )$.
The symmetry constraints for each symmetry class were already taken
into account in the Hamiltonian $\bar{H}$ and $\alpha$, respectively,
so that generically $W(|E|\vec{\epsilon})$ will remain finite in the
limit $|E|\to 0$.  Hence, for energies less than the average level
spacing $\delta_0$, we obtain the asymptotic behavior
\begin{equation}
  \rho(E) \sim |E|^\alpha \qquad |E|\lesssim \delta_0\ .
  \label{eq:alpha}
\end{equation}
The exponents $\alpha$ that result from this argument are listed in
Table~\ref{tab:results} and agree with those stated in
Refs.~\onlinecite{Beenakker2014,Ivanov2002,Altland1997}.  $\alpha$
relates to $q(H)$ in the following way: $\alpha>1$ in classes C and
CII where $q$ cannot be defined. $\alpha=1$ in classes AIII, CI, and
DIII where $q\in\mathbb{C}$.  $\alpha=0$ when $q$ is real.  Finally,
when a matrix ensemble with chiral symmetry is restricted to a subset
of matrices $H$ satisfying $[H,U] = \{C,U\} = 0$, the matrix element
$\braket{\psi}{H_1}{C\psi}$ vanishes and $\alpha$ is reduced to
$\alpha_U = \alpha-1$; cf. Table.~\ref{tab:results}.

\section{Impurity-driven topological phase
  transitions \label{impurity-driven-trans}}
In this section we consider the topological order of a translationally
invariant Hamiltonian $H = H_0 + \lambda H_1'$, where $H_0$ is gapped
and $H_1'$ describes a lattice of impurities with period $\Ni$
determining the overall translational symmetry of the system.  A
general translationally invariant Hamiltonian in dimension $d>0$ is
block diagonal in momentum space,
$H = \bigoplus_{\vec{k}} H (\vec{k})$.\cite{footnote:BZ} Each
Hermitian block $H (\vec{k})$ describes the physics of a $d=0$ system.
In particular, there are $2^d$ high symmetry points $\vec{\Gamma}$ in
the Brillouin zone, which satisfy $\vec{\Gamma}=-\vec{\Gamma}+\vec{G}$
with $\vec{G}$ a reciprocal lattice vector, and each $d=0$ Hamiltonian
$H(\vec{\Gamma})$ belongs to the same symmetry class as $H$
itself. Hence, we can calculate $q(H(\vec{\Gamma}))$ [using the
definition of $q(H)$ for the symmetry class of $H$ itself] to
determine zero-energy crossings of $H(\vec{\Gamma})$.

In the case of the impurity lattice, the blocks in momentum space
become $H_\lambda(\vec{k}) = H_0(\vec{k}) + \lambda H_1'(\vec{k})$.
Finding the zeros $\Lc(\vec{\Gamma})$ of $q(H_\lambda(\vec{\Gamma}))$
yields information about band inversions at $\vec{\Gamma}$ which often
affect the topological invariants of the system.  Let us consider the
case that $H_0 + \Lcinf H_1$ has a zero-energy eigenstate, where $H_1$
is a single impurity of the set of impurities forming $H_1'$.  Its
wave function decays on a length scale $\xi^\infty$ because
$\lambda H_1$ is a local perturbation and $H_0$ is gapped.  Consider
now $H_\lambda(\vec{\Gamma})$ in the limit $\Ni \gg \xi^\infty$.
Then, the boundaries of the unit cell have a vanishing effect such
that
\begin{equation}
  \lim_{\Ni/\xi^\infty\to \infty} \Lc(\vec{\Gamma}) = \Lcinf,
  \label{eq:LimitLc}
\end{equation}
independent of $\vec{\Gamma}$.  However, at finite $\Ni$ the
$\Lc(\vec{\Gamma})$ being unequal give rise to a range of values
$\lambda$ where $H_\lambda$ is nontrivial even though $H_0$ is
trivial.  In systems with sufficiently dilute impurity lattice
(compared to $\Lcinf$), impurity strengths of the order of \Lcinf\ are
nevertheless needed to drive a topological phase transition. Below, we
employ the 1D Kitaev model to analytically illustrate this general
behavior.

The closing of the energy gap can also occur away from the
high-symmetry points of the Brillouin zone. In general, this can
happen at both $\vec{k}$ points distinguished by unitary symmetries of
the Hamiltonian\cite{Slager2013, Fang2012} as well as arbitrary
$\vec{k}$ points, cf. Sec.~\ref{px-pySC}. Then, by considering
$\vec{k}$ as a parameter, one can still use the zero-dimensional
$q(H_\lambda(\vec{k}))$ to analyze the phase transitions. The main
difference is that the zero-dimensional Hamiltonian
$H_\lambda(\vec{k})$, where $\vec{k}$ is considered as a fixed
parameter, usually belongs to a different symmetry class than the full
higher dimensional Hamiltonian containing all the momentum blocks.  We
demonstrate that similar to the case of ferromagnetic Shiba
lattices,\cite{Rontynen2015} it is possible to obtain large Chern
numbers in $p_x+ip_y$ superconductors, but in this case one only needs
nonmagnetic impurities instead of the magnetic atoms. Both of these
models belong to symmetry class D. Moreover, we demonstrate that an
impurity lattice can give rise to large topological invariants in
Chern insulators belonging to symmetry class A.

\subsection{Impurity-driven topological phase transitions in the Kitaev chain}
\label{sec:KitaevChain}

The general ideas presented so far can nicely be illustrated in the
context of the Kitaev model\cite{Kitaev2001} describing a
superconductor of spinless fermions in $d=1$
\begin{equation}
  H_0 = \sum_{i} ( -t c_i^\dagger c_{i+1} + \Delta c_i^\dagger c_{i+1}^\dagger + \mathrm{H.c.}) -\mu \sum_{i} c_i^\dagger c_i. 
  \label{eq:KitaevChainHamiltonian}
\end{equation}
To simplify the analysis we focus on $\Delta = t \in \mathbb{R}$.  The
Hamiltonian $H_0$ belongs to class BDI with artificial TR symmetry
$T=K$ in addition to the intrinsic PH symmetry of Bogoliubov--de
Gennes Hamiltonians.  For $|2t/\mu| < 1$ the system is in a trivial
phase, and for $|2t/\mu| > 1$ the system supports Majorana end modes.
Thus, there is a topological phase transition at $|2t/\mu| =1$.

\begin{figure}
  \includegraphics[width=\columnwidth]{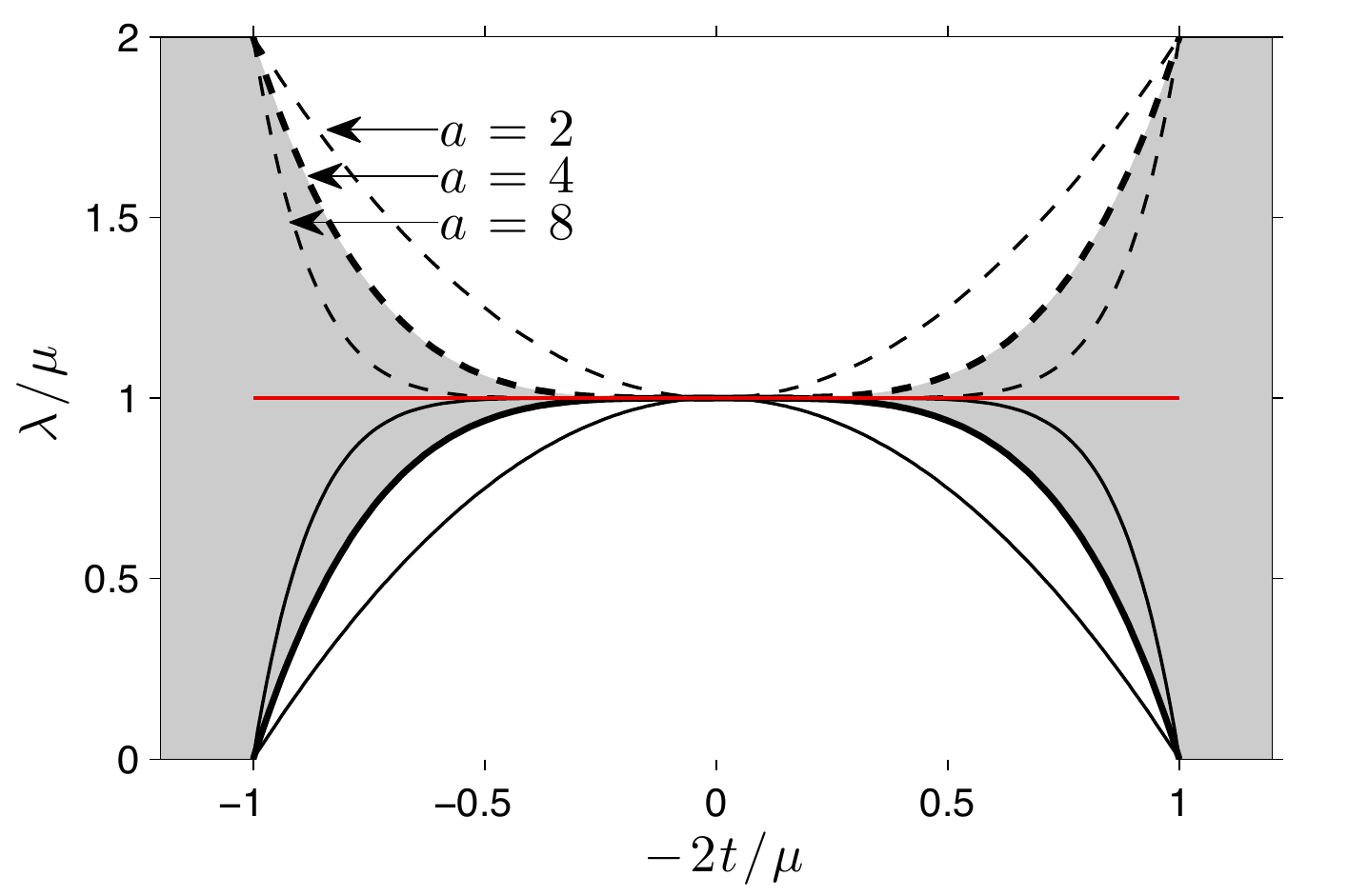}
  \caption{\label{fig:phasediagramKitaev} Phase diagrams of the Kitaev
    chain in the presence of impurity lattices as a function of
    impurity strength $\lambda$ and $2t/\mu$ for $\Delta=t$ and
    different lattice constants $\Ni$.  Gray (white) areas denote
    topologically nontrivial (trivial) phases in case of impurity
    lattice constant $\Ni=4$.  Solid (dashed) lines are $\Lc(0)$
    [$\Lc(\pi)$].  Solid red line is $\Lcinf$.}
\end{figure}

We now consider the effect of an impurity lattice
$\lambda H_1' = \lambda \sum_{i} c_{ai}^\dagger c_{ai}$ on the
topology assuming that we start from inside the topologically trivial
phase $|2t/\mu|<1$.  Here, $\Ni$ is the lattice period.  Band
inversions at the high symmetry $k$ points $\Gamma=0,\pi$ change the
parity of the topological invariant.\cite{Tewari2012} By determining
the zeros of $q(H_\lambda(\Gamma))$, we find that the critical
impurity strengths are (see Appendix~\ref{sec:KitaevChainCalc})
\begin{equation}
  \frac{\Lc(\Gamma)}{\mu} = 1 - e^{i\Gamma} \left(\frac{-2t}{\mu}\right)^{\Ni},
  \qquad \Gamma = 0,\pi
  \label{eq:lcOne}
\end{equation}
with a wave function decaying on the scale
$\xi^\infty = 1/\ln|\mu/2t|$. When $\Ni \gg \xi^\infty$, we get
$\Lc(\vec{\Gamma}) = \Lcinf=\mu$ in agreement with
Eq.~\eqref{eq:LimitLc}.  The phase diagrams for different values of
$\Ni$ are shown in Fig.~\ref{fig:phasediagramKitaev}.  For all values
of $\Ni$, it is possible to turn the trivial phase to a topologically
nontrivial phase with the help of impurity lattice. However for
$\Ni \gg \xi^\infty$ the trivial phase becomes nontrivial only in a
very small region of parameters around $\lambda \approx \Lcinf$.

\begin{figure}
  \includegraphics[width=\columnwidth]{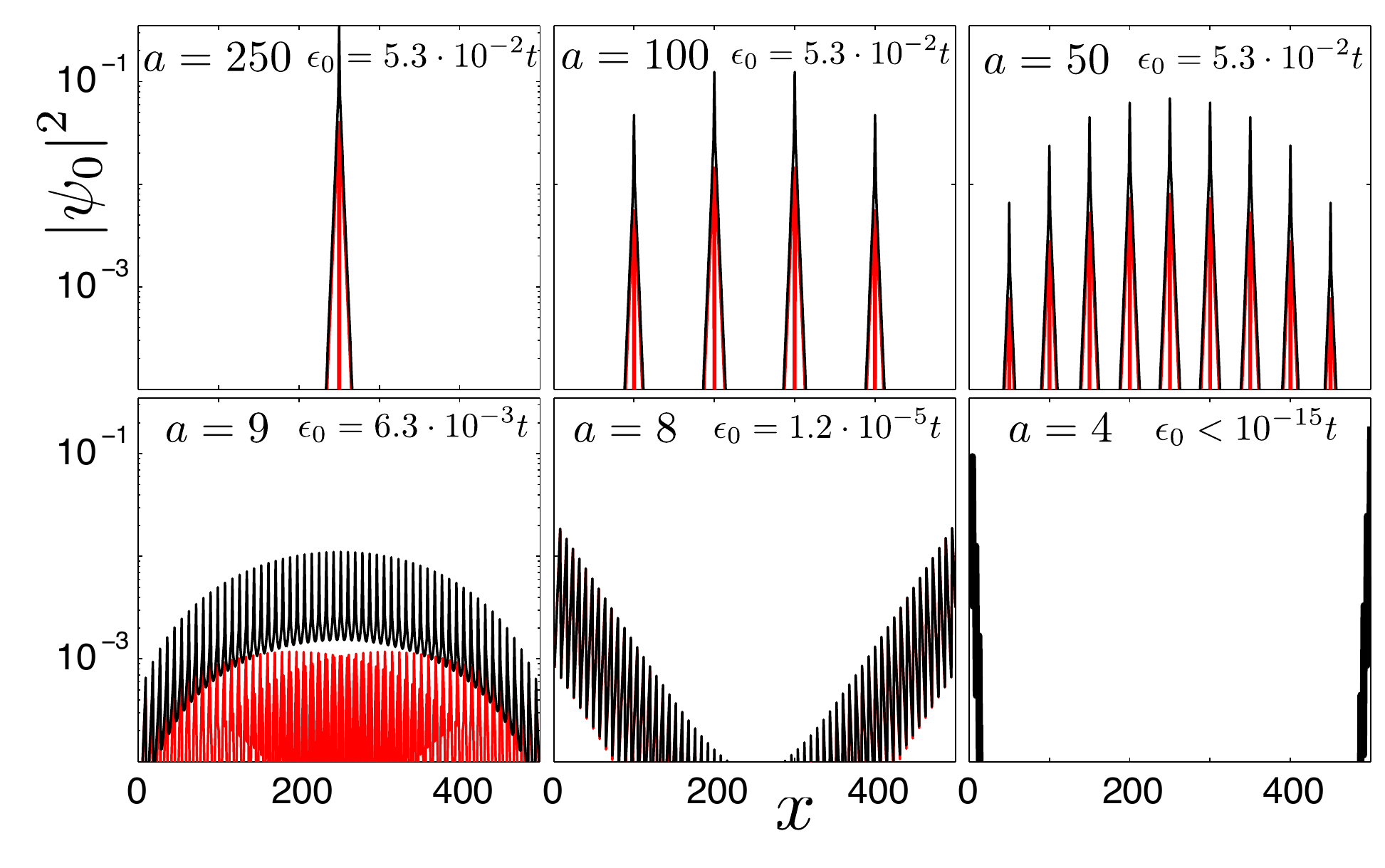}
  \caption{\label{fig:crossover} Illustration of the dimensional
    crossover in the Kitaev chain. A single impurity introduces a
    single subgap state (first panel).  Upon increasing the density of
    impurities (decreasing $a$) the wave functions for the impurity
    states start to overlap and one forms an impurity band.  The
    topology of this band changes and protected end states form, when
    the impurity concentration exceeds its critical value [$a<9$;
    cf. Eq.~\eqref{eq:lcOne}].  $\psi_0$ is the wave function of the
    eigenstate with smallest positive energy $\epsilon_0$.  Black
    (red) curves correspond to the particle (hole) component of
    $\psi_0$.  Parameters: $2t/\mu=0.8$, $\lambda/\mu= 0.85$;
    cf. Fig.~\ref{fig:phasediagramKitaev}.}
\end{figure}

As illustrated in Fig.~\ref{fig:crossover}, our results allow the
following appealing interpretation: A single local impurity can cause
a $d=0$ topological transition.  Taking several $d=0$ systems with an
impurity of strength $\lambda\approx \Lcinf$ and coupling them
together via hopping (and pairing) terms yields a system of
dimensionality $d>0$ with good chances to be topologically different
from the system one would obtain without impurities.

\subsection{Impurity-driven topological phase transitions in
  $p_x + ip_y$ superconductors \label{px-pySC}}

In this section, we consider the two-dimensional $p_x+ip_y$
superconductor described by the Hamiltonian $H_0$ given by
Eq.~\eqref{eq:pxipy}. Since $H_0$ belongs to class D, there exists a
$\mathbb{Z}$-topological invariant, namely the sum $C$ of all Chern
numbers of the occupied bands. The topologically distinct phases of
$H_0$ are $C=+1$ for $0<\mu/t<4$, $C=-1$ for $-4<\mu/t<0$, and $C=0$
for $|\mu/t|>4$.  We will consider this type of superconductor in the
presence of impurities described by the Hamiltonian
$\lambda H_1 = \lambda c_0^\dagger c_0$ arranged on a square lattice
with period $\Ni$ to form a periodic perturbation $\lambda H_1'$.  We
demonstrate that the impurity lattice allows us to access extremely
large Chern numbers similarly as in the case of ferromagnetic Shiba
lattices.\cite{Rontynen2015}

\begin{figure}
  \includegraphics[width=\columnwidth]{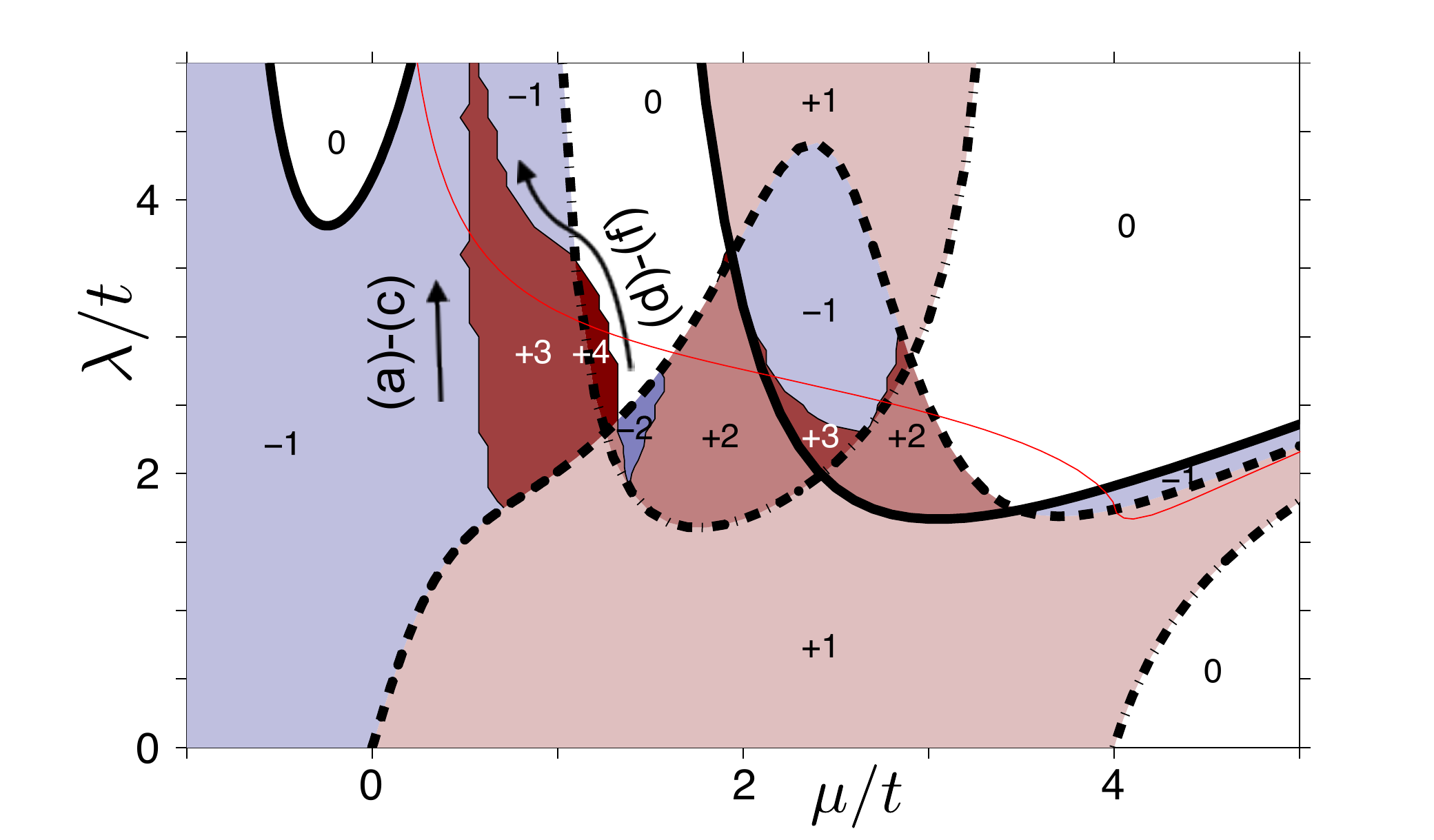}
  \caption{\label{fig:phasediagrampx} Phase diagram of the $p_x+ip_y$
    superconductor in the presence of an impurity lattice as a
    function of impurity strength $\lambda$ and chemical potential
    $\mu$ for $\Delta = 0.5\,t$ and $\Ni = 3$. Thick lines are gap
    closings at the high symmetry points [solid: $(0,0)$, dashed:
    $(0,\pi)$ and $(\pi,0)$, dash-dot: $(\pi,\pi)$].  At thin lines,
    the gap closes away from the high-symmetry points.  Numbers denote
    the sum of Chern numbers of occupied bands in the respective
    phase.  The thin red line is $\Lcinf$. The positions of the gap
    closings in the momentum space for the thin lines denoted as
    (a)-(c) and (d)-(f) are illustrated in
    Fig.~\ref{gapclosingsmomentumspace}.}
\end{figure}

Fig.~\ref{fig:phasediagrampx} shows the phase diagram of the
Hamiltonian $H=H_0+\lambda H_1'$ as a function of $\lambda$ and $\mu$
for $\Delta = 0.5\, t$ and $\Ni = 3$.\cite{footnote:ChernNumbers}
There exists a rich variety of phases, which can be accessed by tuning
these parameters.  The topological phase transitions separating these
phases come in two variants. First, there are transitions accompanied
by a gap closing at the high symmetry points where $k_x=0,\pi$ and
$k_y=0,\pi$.  The corresponding phase transition lines
$\Lc(\vec{\Gamma})$ were obtained from determining the zeros of
$q(H_\lambda(\vec{\Gamma}))$.  At these lines, the Chern number
changes by $\pm 1$.  Notice that the gap closings at $(0,\pi)$ and
$(\pi,0)$ coincide due to rotational symmetry.  Secondly, there are
transitions where the gap closes at four momenta away from the high
symmetry points.  In these cases, the Chern number changes by $\pm
4$. Moreover, the transition lines corresponding to changes of the
Chern number by $\pm 4$ always connect the thick lines corresponding
to gap closings at high symmetry points (see
Fig.~\ref{fig:phasediagrampx}).

\begin{figure}
  \includegraphics[width=\columnwidth]{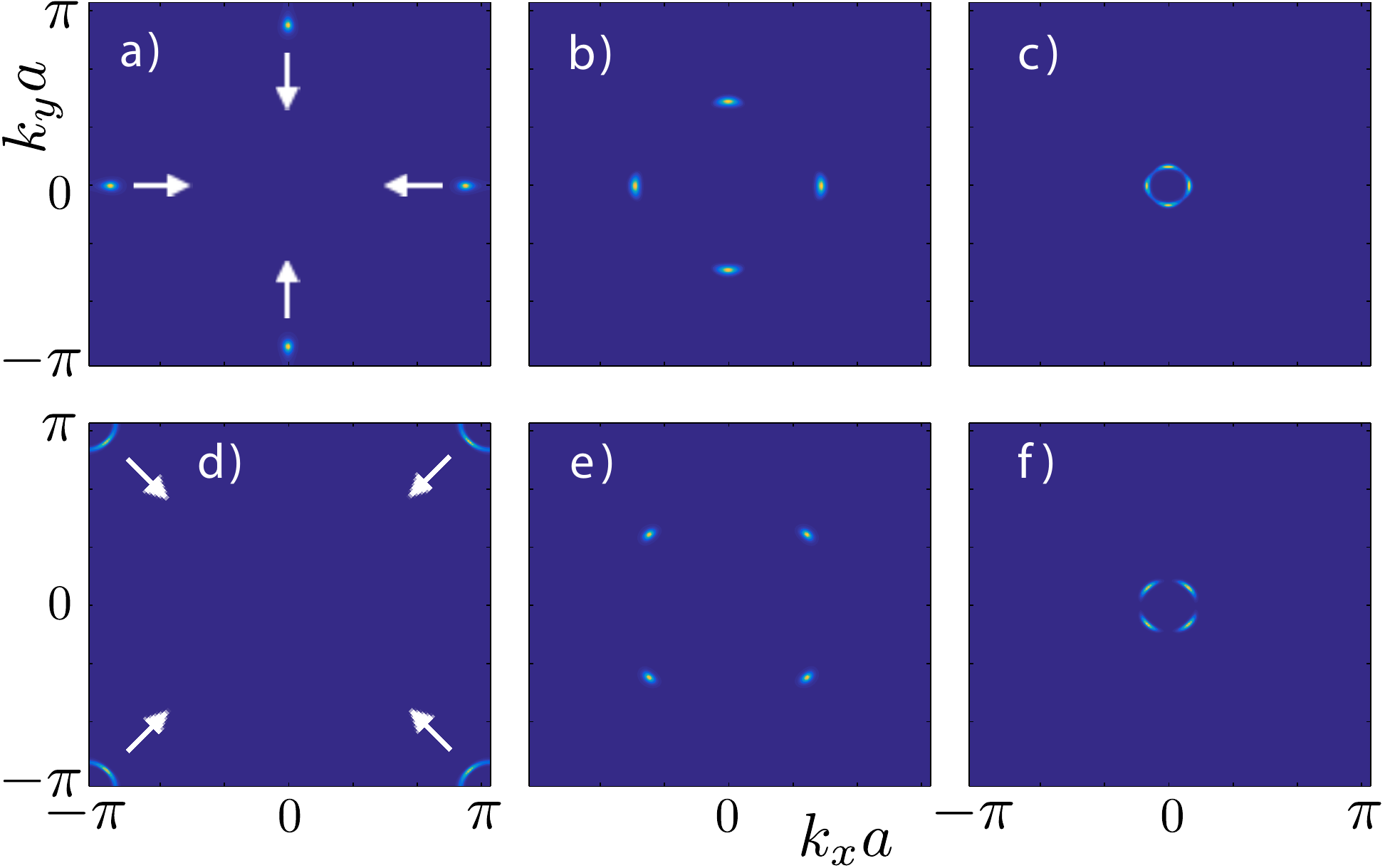}
  \caption{\label{fig:curvatures} Chern curvature of the impurity band
    for different points in the phase diagram in
    Fig.~\ref{fig:phasediagrampx}: (a) $\mu=0.6\,t$,
    $\lambda=1.75\,t$, (b) $\mu=0.5\,t$, $\lambda=4\,t$, (c)
    $\mu=0.48\,t$, $\lambda=6.8\,t$, (d) $\mu=1.37\,t$,
    $\lambda=2\,t$, (e) $\mu=0.85\,t$, $\lambda=4\,t$ and (f)
    $\mu=0.5\,t$, $\lambda=6.8\,t$.  The gap closings associated with
    topological phase transitions give rise to momentum space
    topological defects, which show up as an enhanced Chern curvature
    in the vicinity of the gap closing point in the momentum
    space.} \label{gapclosingsmomentumspace}
\end{figure}

\begin{figure}
  \includegraphics[width=\columnwidth]{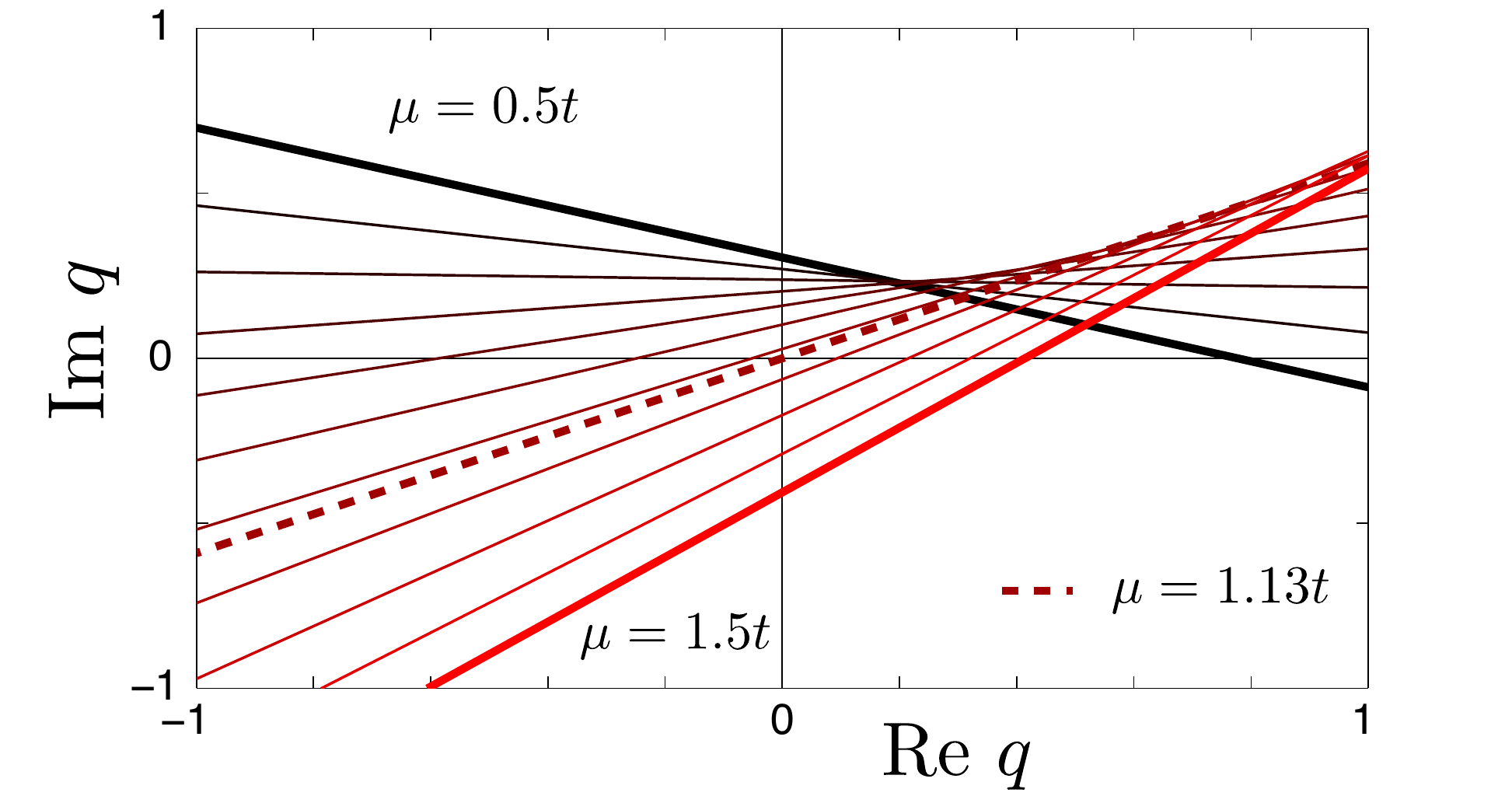}
  \caption{Real and imaginary parts of
    $q(H_\lambda(\vec{k}))=\det D_\lambda(\vec{k})$ as a function of
    $\lambda$ for $k_x a=k_y a=\pi/2$, $\Delta = 0.5\,t$, $\Ni=3$, and
    $\mu \in [0.5\,t, 1.5\,t]$. The lines corresponding to
    $\mu=0.5\,t$ and $\mu=1.5\,t$ are on different sides of the
    origin, so that the Hamiltonian realizes the situation illustrated
    in Fig.~\ref{fig:crossinginplane}. Namely, by tuning
    $\mu \in [0.5\,t, 1.5\,t]$ one smoothly deforms the curve so that
    there exists a particular point $\mu=\mu_c$ and
    $\lambda=\lambda_c$ where $q(H_\lambda(\vec{k}))=0$. This leads to
    a gap closing at $\mu=\mu_c=1.13\,t$ and
    $\lambda=\lambda_c=3.47\,t$ explaining one point in the
    topological phase-transition line denoted (d)-(f) in
    Fig.~\ref{fig:phasediagrampx}. The whole phase-transition line is
    obtained by repeating this analysis for all $k_x=k_y=k$, where
    $ka \in [0, \pi]$. The unitary transformation $V$ block
    off-diagonalizing the Hamiltonian is not unique.  However,
    $q(H_\lambda(\vec{k}))$ is unique up to an overall phase, and
    therefore the different choices of $V$ lead to the same figure up
    to the rotations of the $\Re q$ and $\Im q$ axis.
  } \label{gapclosingsdzero}
\end{figure}

The Hamiltonian obeys reflection symmetries with respect to the
$x$ axis, the $y$ axis, and the two diagonals between these axes.  As
a consequence, $H$ is also fourfold rotationally
symmetric. Therefore, the gap closings corresponding to changes of
Chern number by $\pm 4$ always occur either (i) along the vertical and
horizontal lines connecting $(0,0)$ to $(0, \pm \pi)$ and
$(\pm \pi ,0)$ or (ii) along the diagonal lines connecting $(0,0)$ to
$(\pm \pi, \pm \pi)$ points in the Brillouin zone. The lines denoted
(a)-(c) and (d)-(f) in Fig.~\ref{fig:phasediagrampx} correspond to
these two different situations. To illustrate that this is indeed the
case, we notice that the gap closings associated with topological
phase transitions give rise to momentum space topological defects, so
that we can visualize these topological defects (and gap closings) by
plotting the Chern curvature of the impurity band in the vicinity of
these phase transition lines as shown in Fig.~\ref{fig:curvatures}.

\begin{figure}
  \includegraphics[width=\columnwidth]{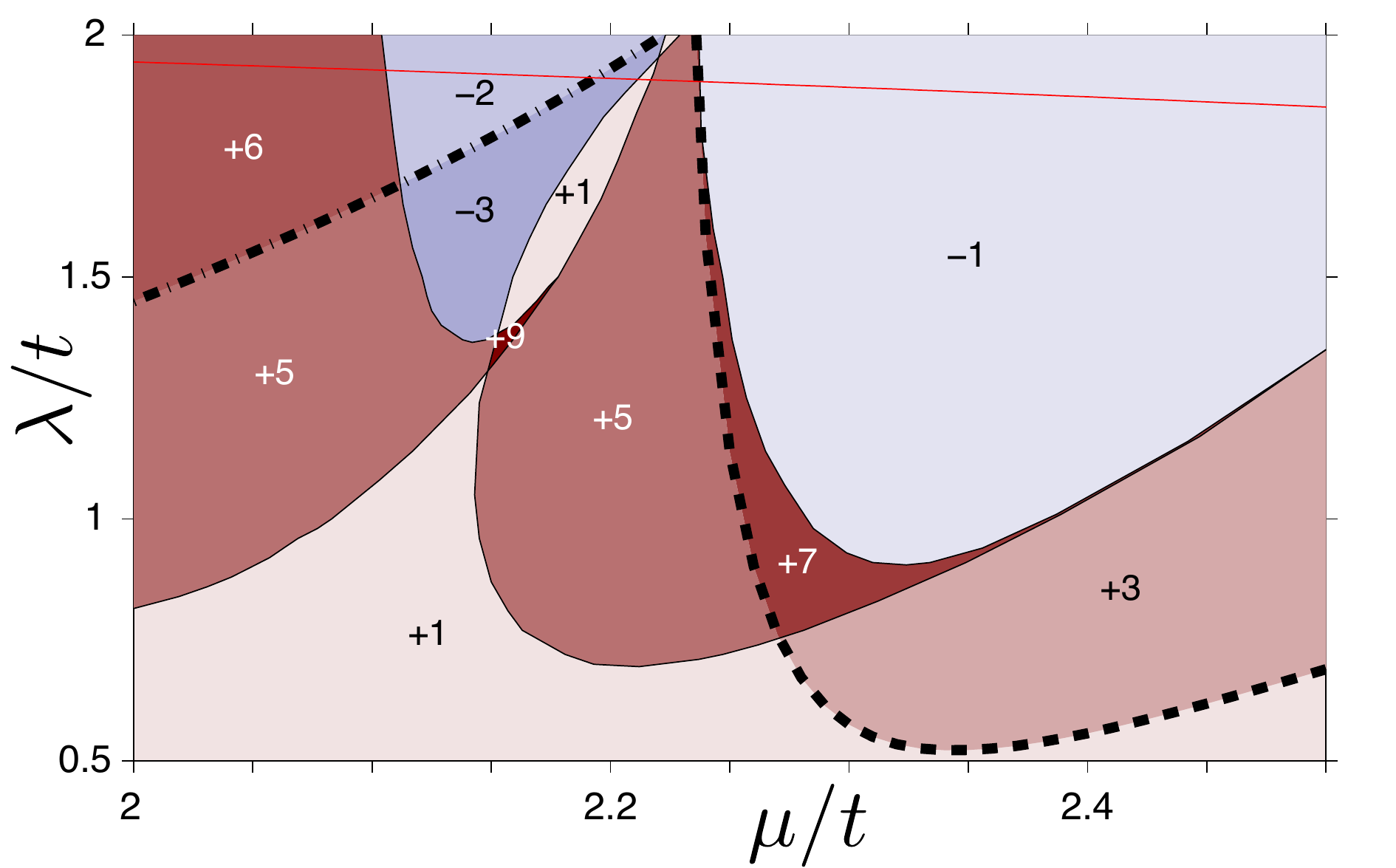}
  \caption{\label{fig:phasediagramlargeC} Same as
    Fig.~\ref{fig:phasediagrampx} but with parameters
    $\Delta = 0.05\,t$ and $\Ni=5$.}
\end{figure}

We can understand also these gap closings occurring away from the
high-symmetry points with the help of the zero dimensional
$q(H_\lambda(\vec{k}))$.  For this purpose we first notice that along
these special lines in the Brillouin zone the zero-dimensional
Hamiltonian $H_\lambda(\vec{k})$ satisfies a chiral symmetry.  The
corresponding chiral symmetry operators can be written as
$C=\tau_y M_x$ for $k_y=0$, $C=\tau_x M_y$ for $k_x=0$,
$C=(\tau_x+\tau_y)M_{xy}/\sqrt{2}$ for $k_x=k_y$, and
$C=(\tau_x-\tau_y)M_{yx}/\sqrt{2}$ for $k_x=-k_y$. Here $M_x$, $M_y$,
$M_{xy}$, and $M_{yx}$ are matrices within the unit cell, which mirror
with respect to the $x$ axis, $y$ axis, $y=x$ line, and $y=-x$ line,
respectively. All these lines are defined in such a way that they run
through the impurity site.  Therefore, as explained in
Sec.~\ref{sec:definingq}, we can use the chiral symmetry to write the
Hamiltonian in the block off-diagonal form
\begin{equation}
  H_\lambda(\vec{k}) = \begin{pmatrix}
    \bm{0} & D_\lambda(\vec{k}) \\ D^\dagger_\lambda(\vec{k}) & \bm{0} 
  \end{pmatrix}
\end{equation}
and define a complex polynomial
$q(H_\lambda(\vec{k})) = \det D_\lambda(\vec{k})$. It is easy to see
that $q(H_\lambda(\vec{k}))$ is always linear in $\lambda$ but a
nonlinear function of $\mu$. Therefore, there can exist some ranges of
$\mu$, where a situation depicted in Fig.~\ref{fig:crossinginplane} is
realized. In Fig.~\ref{gapclosingsdzero}, we plot
$q(H_\lambda(\vec{k}))$ as a function of $\lambda$ for
$k_x a=k_y a=\pi/2$ and $\mu \in [0.5\,t, 1.5\,t]$. The lines
corresponding to $\mu=0.5\,t$ and $\mu=1.5\,t$ are on different sides
of the origin, and by tuning $\mu \in [0.5\,t, 1.5\,t]$ one smoothly
deforms the curve so that there exists a particular point
$\mu=\mu_c=1.13\,t$ and $\lambda=\lambda_c=3.47\,t$ where
$q(H_\lambda(\vec{k}))=0$. These values correspond to one point in the
topological phase transition line denoted (d)-(f) in
Fig.~\ref{fig:phasediagrampx}. The whole phase transition line is
obtained by repeating this analysis for all $k_x=k_y=k$, where
$ka \in [0, \pi]$. Because $q(H_\lambda(\vec{k}))$ is a nonlinear
function of $\mu$ there can exist several ranges of $\mu$ where
similar situations are realized. However, all the phase transition
lines where the Chern number changes by $\pm 4$ (connecting the thick
lines corresponding to gap closings at high symmetry points in
Fig.~\ref{fig:phasediagrampx}) can be explained with a similar
analysis of the parameter values where $q(H_\lambda(\vec{k}))=0$.

\begin{figure}
  \includegraphics[width=\columnwidth]{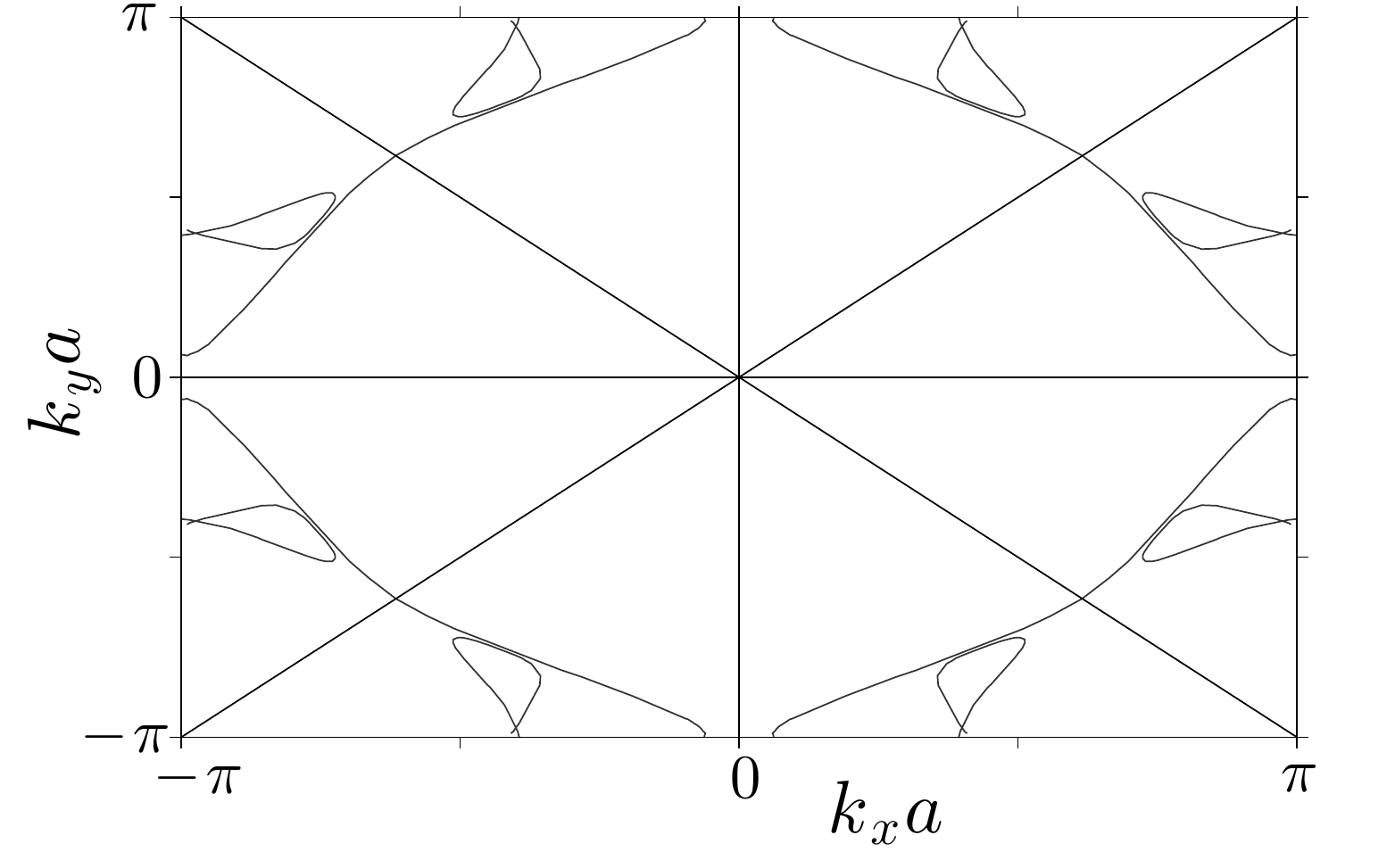}
  \caption{The positions of the energy gap closings in the momentum
    space corresponding to the topological phase transition lines
    shown in Fig.~\ref{fig:phasediagramlargeC} form a structure which
    resembles a spiderweb. The gap closings occurring along the lines
    $k_y=0$, $k_x=0$, $k_x=k_y$, and $k_x=-k_y$ form a support
    structure, and gap closing lines giving rise to changes of Chern
    number by $8$ form curves which begin and end at these support
    lines.} \label{fig:kpositions}
\end{figure}

Finally, we find that even larger Chern numbers can be obtained by
increasing the superconducting coherence length $\xi$ (by decreasing
$\Delta$) and the lattice constant $\Ni$; see
Fig.~\ref{fig:phasediagramlargeC}. This is in agreement with
Ref.~\onlinecite{Rontynen2015}. There, it was argued that Chern
numbers on the order of $\xi/\Ni$ can be obtained in impurity lattices
where $\Ni$ is much larger than the lattice constant. In
Fig.~\ref{fig:phasediagramlargeC}, there are also phase transition
lines where the Chern number changes by $8$. These phase transition
lines always connect the phase transition lines corresponding to gap
closings at the special lines $k_y=0$, $k_x=0$, $k_x=k_y$, and
$k_x=-k_y$. The positions of the corresponding eight gap closings in
the $\vec{k}$ space obey the rotational and reflection symmetries of
the system. Therefore, the positions of the energy gap closings in the
momentum space corresponding to the topological phase transition lines
form a structure, which resembles a spiderweb (see
Fig.~\ref{fig:kpositions}). The gap closings occurring at high
symmetry points and along the lines $k_y=0$, $k_x=0$, $k_x=k_y$, and
$k_x=-k_y$ (possessing chiral symmetries) form the support structure
of this spiderweb, and gap closing points giving rise to changes of
Chern number by $8$ form curves which begin and end at these support
lines.

\subsection{Impurity-driven topological phase transitions in bilayer Chern insulators}

The $p_x+i p_y$ superconductor, similar to the ferromagnetic Shiba
lattice model considered in Ref.~\onlinecite{Rontynen2015}, belongs to
symmetry class D. However, it is clear that the particle-hole symmetry
is not necessary for the appearance of large Chern numbers since small
perturbations breaking this symmetry can not affect the Chern number
due to the presence of the energy gap. In this section, we study a
closely related model, generically not supporting particle-hole
symmetry, which we call bilayer Chern insulator. On one hand, this
model allows us to establish how the breaking of the particle-hole
symmetry affects the phase diagram. On the other hand, it allows us to
generalize the interesting results concerning the impurity lattices
also to topological insulators belonging to symmetry class A.

\begin{figure}
  \includegraphics[width=\columnwidth]{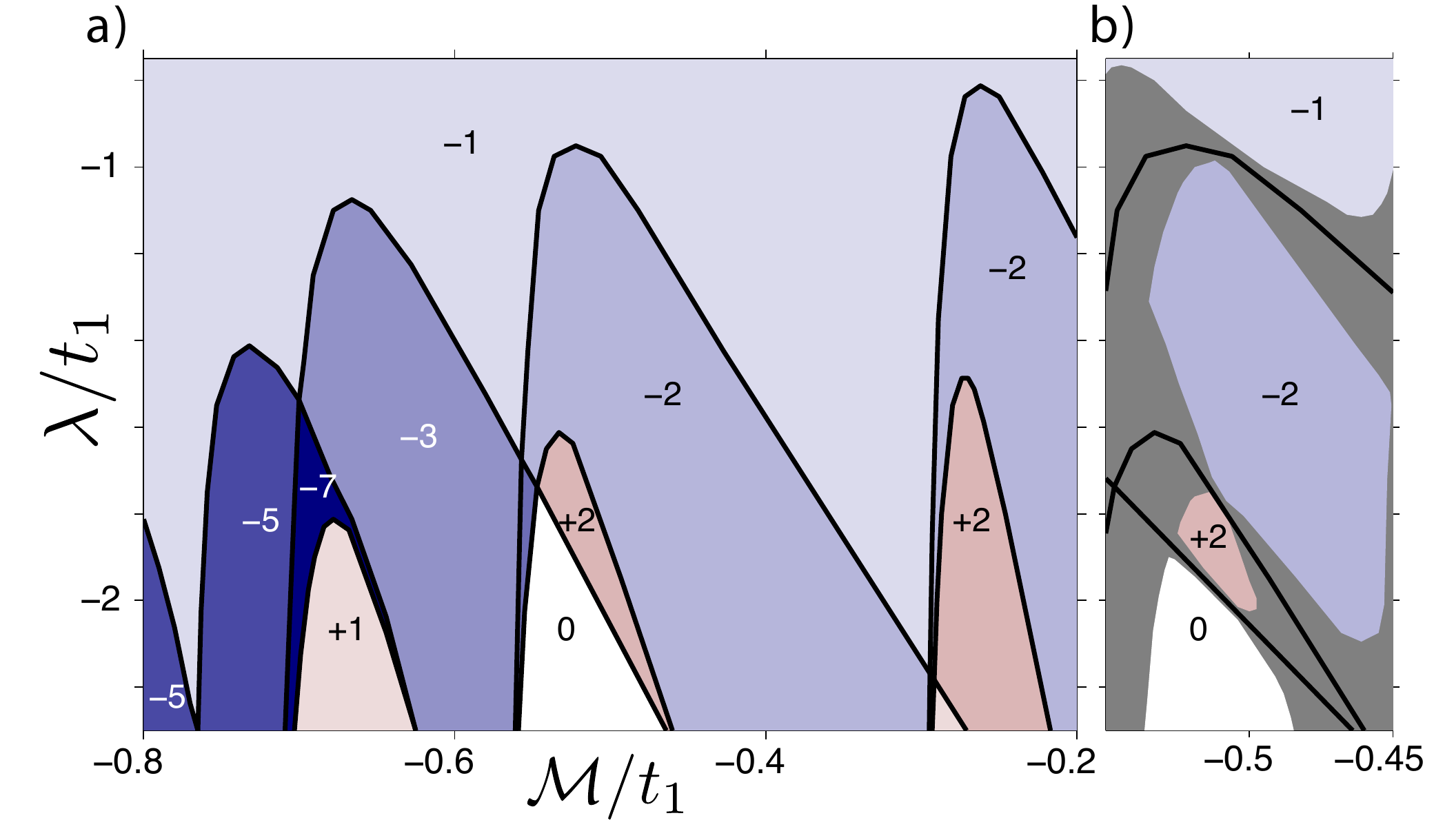}
  \caption{\label{fig:phasediagrambilayer} Phase diagram of the
    bilayer Chern insulator in the presence of an impurity lattice as
    a function of $\lambda$ and ${\cal M}$ for $t_2=0.5\,t_1$,
    ${\cal A}=0.05\,t_1$, $\Ni=7$, and $u=-1/2$. (a) For all values of
    ${\cal M}$ and $\lambda$ the chemical potential $\mu$ is tuned to
    be between the two impurity bands originating from the different
    layers. (b) The chemical potential is fixed to be
    $\mu = 0.17\,t_1$. The gray areas of the parameter space are
    gapless.}
\end{figure}

We assume that the Hamiltonian for a bilayer Chern insulator in the
basis
$c_{\vec{k}}^\dagger=(c_{1,\vec{k}}^\dagger, c_{2,\vec{k}}^\dagger)$
is
\begin{equation}
  H_0(\vec{k}) =
  \begin{pmatrix}
    \xi_1(\vec{k}) & {\cal A} (\sin k_x+i\sin k_y) \\
    {\cal A} (\sin k_x-i\sin k_y) & \xi_2 (\vec{k})
  \end{pmatrix}. 
  \label{eq:Chernbilayer}
\end{equation}
Here $c_{1(2),\vec{k}}^\dagger$ are the electron creation operators
for the different layers,
$\xi_1={\cal M}- 2t_1(\cos k_x + \cos k_y-2)-\mu$ and
$\xi_2=-{\cal M} + 2t_2(\cos k_x + \cos k_y-2)-\mu$ are the
dispersions for the energy bands in the different layers, $t_{1(2)}$
are the hopping amplitudes, $\mu$ is the chemical potential,
${\cal M}$ is the energy gap between bands (${\cal M}<0$ means that
the bands are inverted), and ${\cal A}$ describes the interlayer
tunneling amplitude.  We assume that the energy band in layer 1 is
electronlike with positive effective mass, whereas the energy band in
layer 2 is holelike with negative effective mass. The interlayer
tunneling is assumed to be odd in momentum, which is naturally the
case if the electronlike band is made out of $s$ orbitals and the
holelike band from $p$ orbitals. Thus, our model describes a single
spin block of the BHZ model,\cite{BHZ06} which may in principle be
realized by coupling a quantum spin Hall insulator material to a
ferromagnetic insulator to remove one of the spin
blocks.\cite{footnote:oppositeChern} Furthermore, some of the quantum
spin Hall insulator materials, such as the InAs/GaSb quantum
wells,\cite{Liu08} naturally have the bilayer structure, which will be
important in the following.  Namely, we will consider impurity
lattices, where an impurity on lattice site $\vec{r}$ acts
asymmetrically on the two layers
$\lambda H_1=\lambda (c_{1\vec{r}}^\dagger c_{1\vec{r}}+u
c_{2\vec{r}}^\dagger c_{2\vec{r}})$.  Here, $u$ describes the
asymmetry of the impurity potential, and it can be controlled, for
example, by placing different impurity atoms in the two layers at
position $\vec{r}$. In the following we consider only the case
$u<0$. This gives rise to two impurity bands originating from the two
different layers, and the band inversions occurring between these
impurity bands can cause topological phase transitions. Similarly as
earlier we arrange the impurities on a square lattice with lattice
period $\Ni$.

\begin{figure}
  \includegraphics[width=\columnwidth]{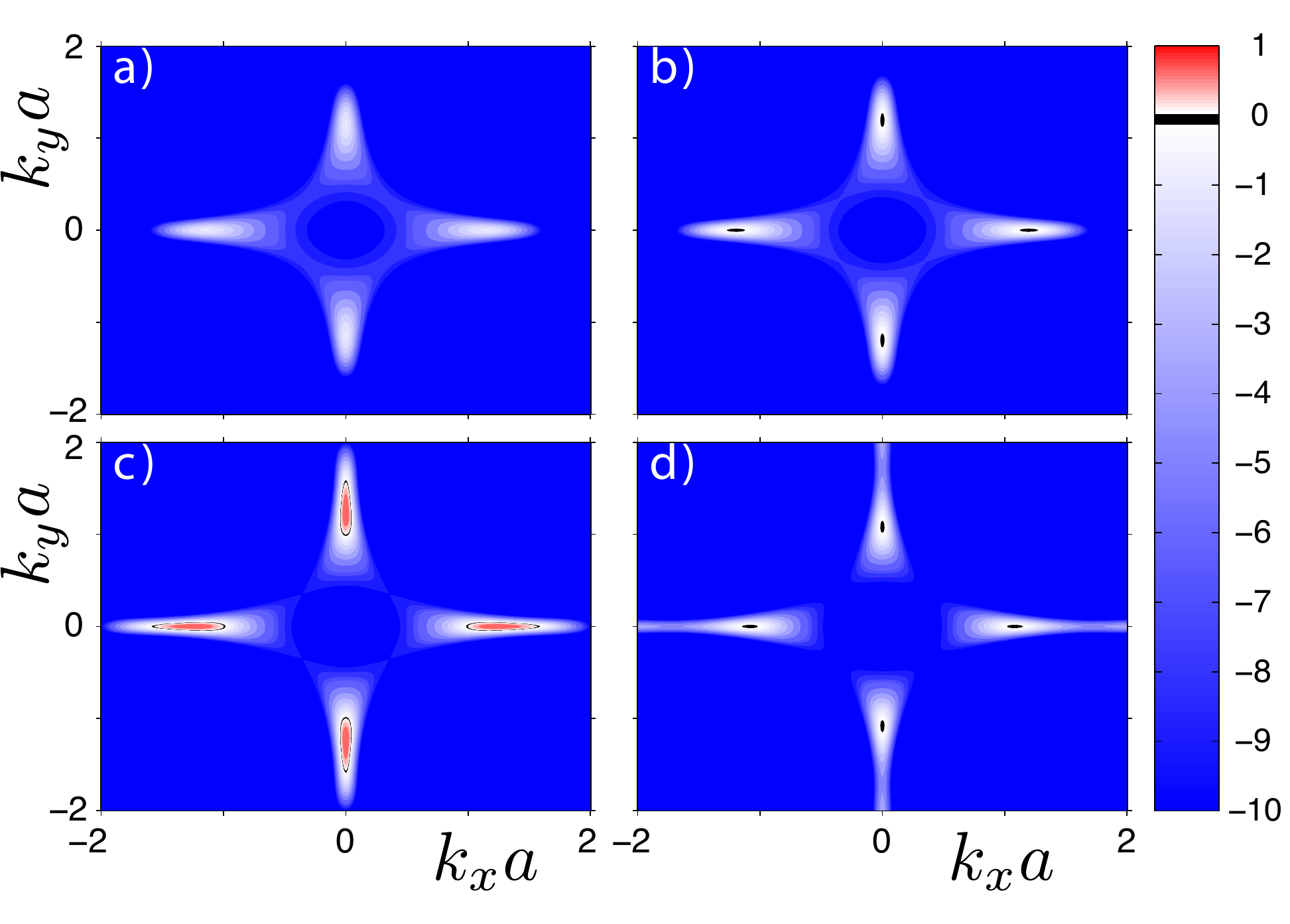}
  \caption{$q(H_\lambda(\vec{k})) = \det H_\lambda(\vec{k})$ for
    ${\cal M} = -0.5\,t_1$ and (a) $\lambda = -1.8t$ (inside $C=-2$
    phase), (b) $\lambda=-1.823 t$ (transition between $C=-2$ phase
    and gapless phase), (c) $\lambda=-1.9 t$ (inside gapless phase),
    (d) $\lambda=-1.96t$ (transition between gapless phase and $C=2$
    phase). The other parameters are the same as in
    Fig.~\ref{fig:phasediagrambilayer}(b).
  } \label{gapclosingsdzerobilayer}
\end{figure}

The phase diagram of the bilayer Chern insulator in the presence of an
impurity lattice as a function of $\lambda$ and ${\cal M}$ is shown in
Fig.~\ref{fig:phasediagrambilayer} for $t_2=0.5\,t_1$,
${\cal A}=0.05\,t_1$, $\Ni=7$, and $u=-1/2$. As can be seen from
Fig.~\ref{fig:phasediagrambilayer}(a), despite the absence of
particle-hole symmetry a rich variety of topologically distinct phases
can be realized also in this model similarly as in the superconducting
model analyzed in Sec.~\ref{px-pySC}. In
Fig.~\ref{fig:phasediagrambilayer}(a), we have tuned the chemical
potential to be between the impurity bands for all values of
${\cal M}$ and $\lambda$ in order to show the overall structure of the
phase diagram for a large range of ${\cal M}$ and $\lambda$
values. However, there are also important differences to the
$p_x+i p_y$ superconductor which become apparent when the chemical
potential remains fixed as in Fig.~\ref{fig:phasediagrambilayer}(b).
Namely, in the absence of particle-hole symmetry, the topologically
distinct phases generically are separated in parameter space by
gapless phases. We can understand this generic feature with the help
of $q(H_\lambda(\vec{k}))$: In symmetry class A, there are no
symmetries present and therefore the zero dimensional invariant is
$q(H_\lambda(\vec{k})) = \det H_\lambda(\vec{k})$. Because
$q(H_\lambda(\vec{k})) = \det H_\lambda(\vec{k})$ is a second order
polynomial in $\lambda$, we generically expect that the topological
phase transitions occur as illustrated for the transition between the
$C=+2$ and $C=-2$ phases in Fig.~\ref{gapclosingsdzerobilayer}.
Inside the fully gapped topological phases, $q(H_\lambda(\vec{k}))$
has the same sign everywhere in momentum space as illustrated for the
$C=+2$ phase in Fig.~\ref{gapclosingsdzerobilayer}(a).  By increasing
$\lambda$ one can reach the transition point between the $C=+2$ phase
and a gapless phase, and at this transition point
$q(H_\lambda(\vec{k})) =0$ at a specific point $\vec{k}_c$ in the
momentum space [Fig.~\ref{gapclosingsdzerobilayer}(b)]. However, by
further increasing $\lambda$, $q(H_\lambda(\vec{k}))$ generically
changes sign in a region around $\vec{k}_c$
[Fig.~\ref{gapclosingsdzerobilayer}(c)] indicating the appearance of
zero-energy states (Fermi surface) at the boundary of this region,
i.e., a gapless phase. When $\lambda$ is further increased, so that
one approaches the $C=-2$ phase, the Fermi surface shrinks again and
finally one reaches a specific value of $\lambda$, where
$q(H_\lambda(\vec{k})) =0$ only at one point in the momentum space
[Fig.~\ref{gapclosingsdzerobilayer}(d)]. This defines the transition
point between the gapless phase and the $C=-2$ phase. We thus find
that generically the phase diagram for bilayer Chern insulator
contains both gapless phases and fully gapped phases with a rich
variety of Chern numbers. In addition to the gapless phases separating
topologically distinct phases, there are also gapless phases
describing semimetal phases with negative indirect gap. The relative
amount of areas with gapless and fully gapped phases in the parameter
space strongly depends on $u$.

\section{Summary and discussion}

In summary, we have identified the general conditions under which the
eigenenergies of the Hamiltonian undergo a robust zero-energy crossing
as a function of external parameters in different Altland-Zirnbauer
symmetry classes. We defined $q(H)$ as a generalized root of $\det H$
and used it to predict or to rule out, respectively, robust
zero-energy crossings in all symmetry classes. We complemented this
result with a perturbation theory analysis, which allows for a
derivation of the asymptotic low-energy behavior of the ensemble
averaged density of states $\rho \sim E^\alpha$ for all symmetry
classes. Finally, we have utilized $q(H(\vec{k}))$ to show that a
lattice of impurities can drive a topologically trivial system into a
nontrivial phase and revealed impurity bands carrying extremely large
Chern numbers in different symmetry classes of two-dimensional
topological insulators and superconductors. This analysis makes it
transparent that $q(H(\vec{k}))$ can be used as a powerful tool in the
analysis of topological phase transitions in higher dimensional
systems.

The impurity bands carrying large Chern number $C$ are interesting on
their own. In the bilayer Chern insulator, they will support a large
number ($|C|$) of topologically protected fermionic edge modes
allowing dissipationless transport, and in topological superconductors
there will be $|C|$ Majorana edge modes. However, we also want to
point out that the bandwidth of the impurity bands can be quite small,
and it may be tuned by varying the distance between impurities.
Therefore, these impurity bands carrying large Chern numbers can
potentially also support interesting correlated phases in the presence
of interactions. In particular, the interactions may lead to
spontaneous symmetry breaking, such as superconductivity,
ferromagnetism, or exciton condensation, and the physical properties,
such as the charge of topological defects and superfluid density, are
expected to depend on $C$.\cite{Sondhi93,Moon95,Pikulin15,Torma15}
Moreover, partially filled nearly flat bands with a large Chern number
in the presence of interactions may lead to exotic strongly correlated
states of matter with interesting topological
properties.\cite{Barkeshli12, Bergholtz12, Bernevig13}

\begin{acknowledgments}
  L.K. acknowledges valuable discussions with H.-G.~Zirnstein.  This
  work was supported by DFG Grant RO 2247/8-1, by the Academy of
  Finland through its Center of Excellence program, and by the
  European Research Council (Grant No.~240362-Heattronics).
\end{acknowledgments}

\appendix

\section{Topological invariants in the presence of an additional
  symmetry $U$ anticommuting with $C$}
\label{sec:Implications}
In this section, we derive the block diagonal structure for $H$
satisfying a chiral symmetry in the presence of an additional symmetry
$[H,U]=\{C,U\}=0$, with $U^\dagger U = 1$.  Symmetries of that kind
have been considered before for class DIII in
Refs.~\onlinecite{Woelms2014, Kimme2015}.  The complementary case
where $U$ and $C$ commute was studied in
Ref.~\onlinecite{Koshino2014}.

The simultaneous eigenstates of $H$ and $U$, $H\ket{E,u}=E\ket{E,u}$
and $U\ket{E,u}=u\ket{E,u}$ satisfy
\begin{equation}
  C\ket{E,u} = \ket{-E,-u}. \label{eq:mappingChiralPartners}
\end{equation}
Choosing the states $\ket{E,u}$ as a basis of the Hilbert space and
sorting them appropriately, the operators $U$, $H$, and $C$ become
block diagonal:
\begin{subequations}
\begin{align}
  U &= \bigoplus_{i=1}^{n} \begin{pmatrix}
    u_i \bm{1}_{m_i} & \bm{0}_{m_i}  \\
    \bm{0}_{m_i} & -u_i \bm{1}_{m_i}
  \end{pmatrix}, \\
  H &= \bigoplus_{i=1}^{n} \begin{pmatrix}
    H_{+i} & \bm{0}_{m_i} \\
    \bm{0}_{m_i} & H_{-i}
    \end{pmatrix}, \label{eq:Hblocks}\\
  C &= \bigoplus_{i=1}^{n} \begin{pmatrix}
    \bm{0}_{m_i} & C_i \\
    C_i^\dagger & \bm{0}_{m_i}
  \end{pmatrix}.
\end{align}
\end{subequations}
Here, $m_i$ denotes the degree of degeneracy of the eigenvalue $u_i$.
The specific form of $C$ follows from
Eq.~\eqref{eq:mappingChiralPartners} and the property
$C^\dagger C = C^2 = 1$.  Each of the Hamiltonian's blocks $H_{\pm i}$
lacks chiral symmetry, although the full Hamiltonian still has chiral
symmetry which requires $H_{-i} = - C_i^\dagger H_{+i} C_i$.
Therefore the number of SPEs per block is halved from the original
value $\nSPE_{\pm i} = \nu / 2$.  Consequently, the blocks $H_{\pm i}$
belong to a different symmetry class than $H$, namely to (i) A when
$H$ belongs to AIII, (ii) A or AI when $H$ belongs to CI, (iii) A or
AI when $H$ belongs to BDI, (iv) D or AII when $H$ belongs to DIII,
and (v) C or AII when $H$ belongs to CII.  So in classes AIII, CI, and
DIII the symmetry $U$ prevents level repulsion and enables the
definition of an invariant
${\cal Q}_U = ({\cal Q}_{H_{+1}}, \dots, {\cal Q}_{H_{+n}})$, which is
the vector of invariants of independent blocks $H_{+i}$.  This is not
necessarily the case when $H$ is from CII, since in this case all
blocks can belong to class C.

\section{Impossibility of defining $q(H)$ in classes C and CII}
\label{sec:classesCandCII}
In symmetry classes C and CII, it is not possible to calculate the
generalized second and fourth root of $\det H$, respectively.  To show
this, it suffices to demonstrate the impossibility of defining a
polynomial expression $q(H)$ in the matrix elements of a Hamiltonian
from class C, which would satisfy $|\det H| = |q(H)|^2$.  From this,
the impossibility of defining $q(H)$ for arbitrary Hamiltonians in
class CII follows, because a simple Hamiltonian in CII can, for
example, be the direct sum
$H_\mathrm{CII}^0 = H_\mathrm{C} \oplus H_\mathrm{C}'$ of two
Hamiltonians from class C which are related to each other by TR,
$H_\mathrm{C}' = T H_\mathrm{C} T^{-1}$ with $T^2=-1$.

The simplest Hamiltonian in class C reads
$H_\mathrm{C}^0 =
-\vec{B}\cdot\vec{\sigma}$.\cite{footnote:HCsymmetry}
Here, $\vec{B} = (\Re b_\perp,\Im b_\perp,b_\parallel)$ and
$\vec{\sigma} = (\sigma_1,\sigma_2,\sigma_3)$ is the vector of Pauli
matrices.  Considering
\begin{equation}
  |\det H_\mathrm{C}^0| = \left| b_\parallel + i|b_\perp|  \right|^2,
\end{equation}
it becomes manifest that any candidate for $q(H)$ in class C would
involve the calculation of absolute values or square roots, like
$|b_\perp|$ or $\sqrt{b_\perp^* b_\perp}$, respectively.  Hence, there
is no expression $q(H)$ in class C that meets the criterion of being a
polynomial expression in the Hamiltonian's matrix elements.

\section{Calculations for the Kitaev Chain}
\label{sec:KitaevChainCalc}
This section contains details about the derivation of the results for
the Kitaev chain with impurity lattice in Sec.~\ref{sec:KitaevChain}.
We consider the translationally invariant Hamiltonian
$H_\lambda = H_0 + \lambda H_1'$ with $H_0$ given by
Eq.~\eqref{eq:KitaevChainHamiltonian} and
$\lambda H_1' = \lambda \sum_{i} c_{ai}^\dagger c_{ai}$.  For the
Kitaev chain, bringing $H_\lambda$ into the block off-diagonal form of
Eq.~\eqref{eq:HamiltonianBlockoffdiagonal} corresponds to transforming
to the Majorana basis
$(\gamma_{A,1}, \dots, \gamma_{A,N}, \gamma_{B,1}, \dots,
\gamma_{B,N})$
where $\gamma_{A,j} = \frac{1}{\sqrt{2}}(c_j+c_j^\dagger)$ and
$\gamma_{B,j} =
\frac{i}{\sqrt{2}}(c_j-c_j^\dagger)$.\cite{Beenakker2014}
In momentum space, the off-diagonal block $D_\lambda(\Gamma)$ at the
high symmetry points $\Gamma=0,\pi$ reads
\begin{equation}
  D_\lambda(\Gamma) = -i
  \begin{pmatrix}
    -\mu+\lambda & -2t & &  \\
    0 & -\mu & \ddots &\\
    & \ddots & \ddots & -2t\\
    -2te^{i\Gamma}& & 0 & -\mu
  \end{pmatrix}
\end{equation}
with matrix dimension $a$.  It is then straightforward to determine
the critical impurity strengths [Eq.~\eqref{eq:lcOne}] from the
condition $\det D_\Lc=0$.  Moreover, the corresponding normalized
eigenvector $\psi$ satisfying $D_\Lc \psi = 0$ is
\begin{equation}
  \psi = \sqrt{\frac{\beta^2-1}{\beta^{2\Ni}-1}} (e^{-i\Gamma}, \beta^{\Ni-1}, \beta^{\Ni-2},\dots, \beta)^T, 
\end{equation}
where $\beta = -2t/\mu$.  From these eigenvectors, one obtains the
decay length $\xi^\infty = 1/\ln|\mu/2t|$.


\begin{thebibliography}{99}
\bibitem{footnote:ChiralSymmetry} In the case of
  $\tr C = n \in \mathbb{Z}\setminus\{0\}$, the present discussion is
  not of interest anyway, because there will be $|n|$ zero-energy
  states protected by the chiral symmetry, so that $\det H \equiv 0$.
\bibitem{footnote:BZ} Except for the definitions of the Hamiltonians
  $H_0$ in Eqs.~\eqref{eq:pxipy} and \eqref{eq:Chernbilayer}, which do
  not contain an impurity lattice, all discussion in
  Sec.~\ref{impurity-driven-trans} refers to the reduced Brillouin
  zone in the presence of an impurity lattice.
\bibitem{footnote:ChernNumbers} The Chern numbers in
  Figs.~\ref{fig:phasediagrampx}, \ref{fig:phasediagramlargeC}, and
  \ref{fig:phasediagrambilayer} were obtained by connecting points in
  the gapped phases with $\lambda\neq 0$ to points at the lines
  $\lambda=0$ and calculating the change $\Delta C_\mathrm{imp}$ of
  the impurity band's Chern number at each gap closing line.
\bibitem{footnote:oppositeChern} Without the ferromagnetic insulator
  to break time-reversal symmetry, the physics discussed here can
  still be realized, but then the two spin blocks will give rise to
  opposite Chern numbers.
\end{thebibliography}
\end{document}